# On the Error Exponents of Some Improved Tangential-Sphere Bounds

Moshe Twitto Igal Sason

Department of Electrical Engineering
Technion – Israel Institute of Technology
Haifa 32000, Israel
{tmoshe@tx, sason@ee}.technion.ac.il

June 28, 2018

**Abstract**

The performance of maximum-likelihood (ML) decoded binary linear block codes over the AWGN channel is addressed via the tangential-sphere bound (TSB) and two of its recent improved versions. The paper is focused on the derivation of the error exponents of these bounds. Although it was exemplified that some recent improvements of the TSB tighten this bound for finite-length codes, it is demonstrated in this paper that their error exponents coincide. For an arbitrary ensemble of binary linear block codes, the common value of these error exponents is explicitly expressed in terms of the asymptotic growth rate of the average distance spectrum.

## 1 Introduction

In recent years, much effort has been put into the derivation of tight performance bounds on the error probability of linear block codes under soft-decision maximum-likelihood (ML) decoding. During the last decade, this research work was stimulated by the introduction of various codes defined on graphs and iterative decoding algorithms, achieving reliable communication at rates close to capacity with feasible complexity. The remarkable performance of these codes at rates remarkably above the cut-off rate, makes the union bound useless for their performance evaluation. Hence, tighter performance bounds are required to gain some insight on the performance of these efficient codes. Improved upper and lower bounds on the error probability of linear codes under ML decoding are addressed in [12] and references therein, and applied to various codes and ensembles.

The tangential-sphere bound (TSB) [9] forms one of the tightest performance bounds for ML decoded linear block codes whose transmission takes place over the binary-input additive white Gaussian noise (BIAWGN) channel. The TSB was modified by Sason and Shamai [10] for the analysis of the bit error probability of linear block codes, and was slightly refined by Zangl and Herzog [19]. This bound only depends on the distance spectrum of the code (or the input-output weight enumerating function (IOWEF) of the code for the bit-error analysis [10]), and hence, it can be applied to various codes or ensembles. The TSB falls within the class of upper bounds whose derivation relies on the basic inequality

$$\Pr(\text{word error} \mid \mathbf{c}_0) \leq \Pr(\text{word error},\ \mathbf{y} \in \mathcal{R} \mid \mathbf{c}_0) + \Pr(\mathbf{y} \notin \mathcal{R} \mid \mathbf{c}_0) \qquad (1)$$

where $\mathbf{c}_0$ is the transmitted codeword, $\mathbf{y}$ denotes the received vector at the output of the channel, and $\mathcal{R}$ designates an arbitrary geometrical region which can be interpreted as a subset of the observation space. The basic idea of this bounding technique is to reduce the number of overlaps between the decision regions associated with the pairwise error probabilities used for the calculation of union bounds. This is done by separately bounding the error events for which the noise resides in a region $\mathcal{R}$. The TSB, for example, uses a circular hyper-cone as the region $\mathcal{R}$. Other upper bounds from this family are addressed in [12, Sections 3 and 4], [18] and references therein. In [15], Yousefi and Khandani prove that among all the volumes $\mathcal{R}$ which possess some symmetry properties, the circular hyper-cone yields the tightest bound. This finding demonstrates the optimality of the TSB among a family of bounds associated with geometrical regions which possess some symmetry properties, and which are obtained by applying the *union bound* on the first term in the RHS of (1). In [16], Yousefi and Khandani suggest to use the Hunter bound [8] (an upper bound which belongs to the family of second-order Bonferroni-type inequalities) instead of the union bound. This modification should result in a tighter upper bound, and they refer to the resulting upper bound as the added hyper plane (AHP) bound. Yousefi and Mehrabian also apply the Hunter bound, but implement it in a quite different way in order to obtain an improved tangential-sphere bound (ITSB) which solely depends on the distance spectrum of the code. The tightness of the ITSB and the AHP bound is exemplified in [16, 17] for some short linear block codes, where these bounds slightly outperform the TSB at the low SNR range.

An issue which is not addressed analytically in [16, 17] is whether the new upper bounds (namely, the AHP and the ITSB) provide an improved lower bound on the error exponent as compared to the error exponent of the TSB. In this paper, we address this question, and prove that the error exponents of these improved tangential-sphere bounds coincide with the error exponent of the TSB. We note however that the TSB fails to reproduce the random coding error exponent, especially for high-rate linear block codes [9].

This paper is organized as follows: The TSB ([9], [10]), the AHP bound [16] and the ITSB [17] are presented as a preliminary material in Section 2. In Section 3, we derive the error exponents of the ITSB and the AHP, respectively and state our main result. We conclude our discussion in Section 4. An Appendix provides supplementary details related to the proof of our main result.

## 2 Preliminaries

We introduce in this section some preliminary material which serves as a preparatory step towards the presentation of the material in the following section. We also present notation from [1] which is useful for our analysis. The reader is referred to [12, 18] which introduce material covered in this section. However, in the following presentation, we consider boundary effects which were not taken into account in the original derivation of the two improved versions of the TSB in [16]–[18]). Though these boundary effects do not have any implication in the asymptotic case where we let the block length tend to infinity, they are addressed in this section for finite block lengths.

## 2.1 Assumption

Throughout this paper, we assume a binary-input additive white Gaussian noise (AWGN) channel with double-sided power spectral density of $\frac{N_0}{2}$. The modulation of the transmitted signals is antipodal, and the modulated signals are coherently detected and ML decoded (with soft decision).

## 2.2 Tangential-Sphere Bound

The TSB forms an upper bound on the decoding error probability of ML decoding of linear block code whose transmission takes place over a binary-input AWGN channel [9, 10]. Consider an $(n, k)$ linear block code $\mathcal{C}$ of rate $R \triangleq \frac{k}{n}$ bits per channel use. Let us designate the codewords of $\mathcal{C}$ by $\{\mathbf{c}_i\}$, where $i = 0, 1, \ldots, 2^k - 1$. Assume a BPSK modulation and let $\mathbf{s}_i \in \{+\sqrt{E_\mathrm{s}}, -\sqrt{E_\mathrm{s}}\}^n$ designate the corresponding equi-energy, modulated vectors, where $E_\mathrm{s}$ designates the transmitted symbol energy. The transmitted vectors $\{\mathbf{s}_i\}$ are obtained from the codewords $\{\mathbf{c}_i\}$ by applying the mapping $\mathbf{s}_i = (2\mathbf{c}_i - \mathbf{1})\sqrt{E_\mathrm{s}}$, so their energy is $nE_\mathrm{s}$. Since the channel is memoryless, the received vector $\mathbf{y} = (y_1, y_2, \ldots, y_n)$, given that $\mathbf{s}_i$ is transmitted, can be expressed as

$$y_j = s_{i,j} + z_j \ , \quad j = 1, 2, \ldots, n \tag{2}$$

where $s_{i,j}$ is the $j^{\text{th}}$ component of the transmitted vector $\mathbf{s}_i$, and $\mathbf{z} = (z_1, z_2, \ldots, z_n)$ designates an $n$-dimensional Gaussian noise vector which corresponds to $n$ orthogonal projections of the AWGN. Since $\mathbf{z}$ is a Gaussian vector and all its components are un-correlated, then the $n$ components of $\mathbf{z}$ are i.i.d., and each component has a zero mean and variance $\sigma^2 = \frac{N_0}{2}$.

Let $E$ be the event of deciding erroneously (under ML decoding) on a codeword other than the transmitted codeword. The TSB is based on the central inequality

$$\Pr(E|\mathbf{c}_0) \leq \Pr(E, \mathbf{y} \in \mathcal{R}|\mathbf{c}_0) + \Pr(\mathbf{y} \notin \mathcal{R}|\mathbf{c}_0) \tag{3}$$

where $\mathcal{R}$ is an $n$-dimensional circular cone with a half angle $\theta$ and a radius $r$, whose vertex is located at the origin and whose main axis passes through the origin and the point corresponding to the transmitted vector (see Fig.1). The optimization is carried over $r$ ($r$ and $\theta$ are related as shown in Fig. 1). Let us designate this circular cone by $C_n(\theta)$. Since we deal with linear codes, the conditional error probability under ML decoding does not depend on the transmitted codeword of the code $\mathcal{C}$, so without any loss of generality, one can assume that the all-zero codeword, $\mathbf{s}_0$, is transmitted. Let $z_1$ be the radial component of the noise vector $\mathbf{z}$ (see Fig. 1) so the other $n-1$ components of $\mathbf{z}$ are orthogonal to the radial component $z_1$. From Fig. 1, we obtain that

$$\begin{aligned} r &= \sqrt{nE_\mathrm{s}} \tan\theta \\ r_{z_1} &= \left(\sqrt{nE_\mathrm{s}} - z_1\right) \tan\theta \\ \beta_k(z_1) &= \left(\sqrt{nE_\mathrm{s}} - z_1\right) \tan\zeta = \frac{\sqrt{nE_\mathrm{s}} - z_1}{\sqrt{nE_\mathrm{s} - \frac{\delta_k^2}{4}}} \frac{\delta_k}{2} \end{aligned} \tag{4}$$

The random variable $Y \triangleq \sum_{i=2}^{n} z_i^2$ is $\chi^2$ distributed with $n-1$ degrees of freedom, so its *pdf* is given by

$$f_Y(y) = \frac{y^{\frac{n-3}{2}} e^{-\frac{y}{2\sigma^2}} U(y)}{2^{\frac{n-1}{2}} \sigma^{n-1} \Gamma\left(\frac{n-1}{2}\right)}, \quad y \geq 0 \tag{5}$$

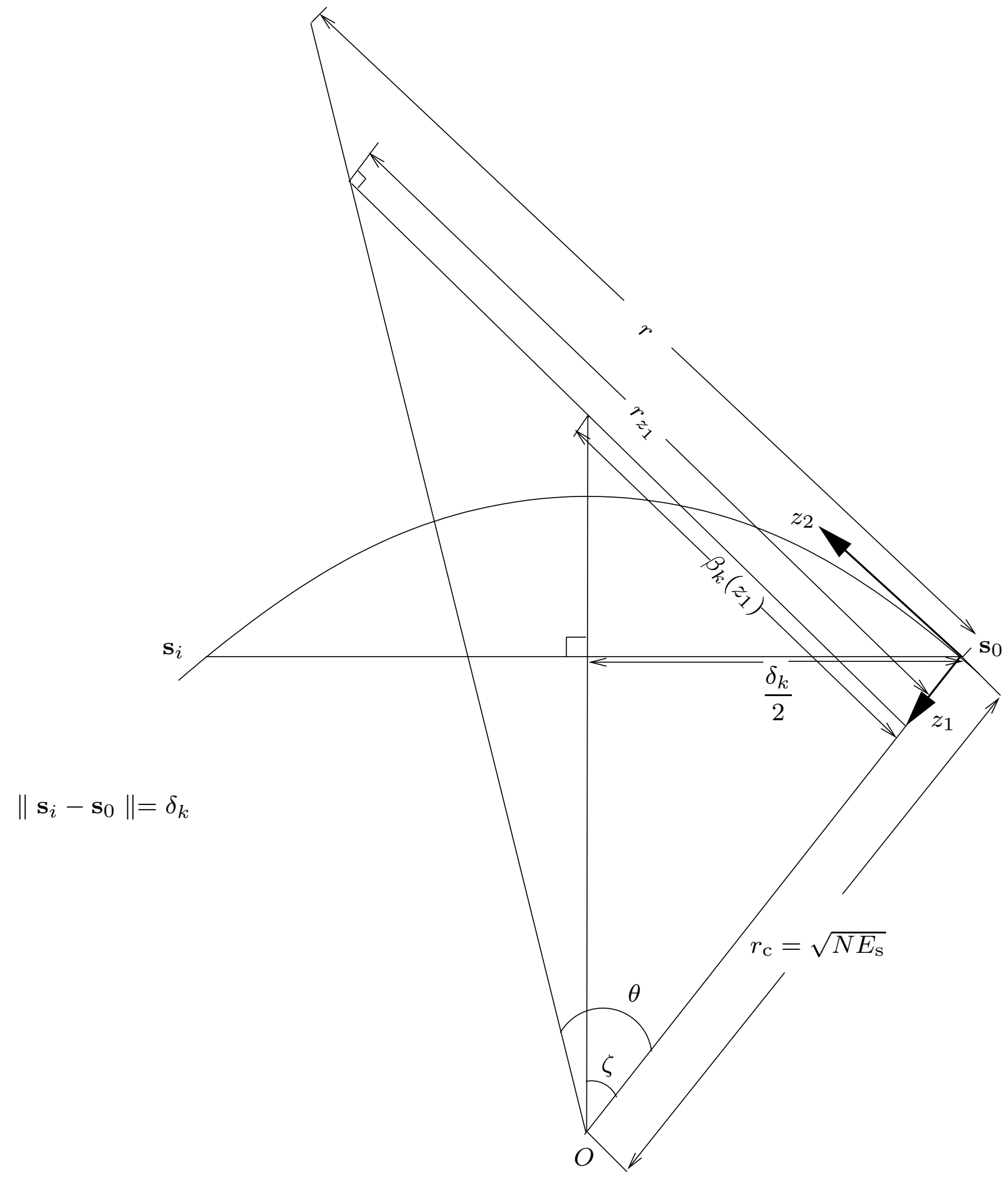

Figure 1: The geometric interpretation of the TSB.

where $U$ designates the unit step function, and the function $\Gamma$ is the complete Gamma function

$$\Gamma(x) = \int_0^\infty t^{x-1} e^{-t} dt, \quad \text{Real}(x) > 0. \tag{6}$$

Conditioned on the value of the radial component of the noise, $z_1$, let $E(z_1)$ designate the decoding error event. The conditional error probability satisfies the inequality

$$\Pr(E(z_1) \mid z_1) \leq \Pr\left(E(z_1), \mathbf{y} \in C_n(\theta) \mid z_1\right) + \Pr\left(\mathbf{y} \notin C_n(\theta) \mid z_1\right) \tag{7}$$

The conditional error event $E(z_1)$ can be expressed as a union of pairwise error events, so

$$\Pr(E(z_1), \mathbf{y} \in C_n(\theta) \mid z_1) = \Pr\left(\bigcup_{i=1}^{M-1} E_{0 \to i}(z_1), \mathbf{y} \in C_n(\theta) \mid z_1\right), \qquad M \triangleq 2^k \tag{8}$$

where $E_{0\to i}(z_1)$ designates the event of error had the only codewords been $\mathbf{c}_0$ and $\mathbf{c}_i$, given the value $z_1$ of the radial component noise in Fig. 1, and $M \triangleq 2^k$ denotes the number of codewords of the code $\mathcal{C}$. We note that for BPSK modulation, the Euclidean distance between the two signals $\mathbf{s}_i$ and $\mathbf{s}_0$ is directly linked to the Hamming weight of the codeword $\mathbf{c}_i$. Let the Hamming weight of $\mathbf{c}_i$ be $h$, then the Euclidean distance between $\mathbf{s}_0$ and $\mathbf{s}_i$ is equal to $\delta_h = 2\sqrt{hE_\mathrm{s}}$. Let $\{A_h\}$ be the distance spectrum of the linear code $\mathcal{C}$, and let $E_h(z_1)$ be the event of deciding under ML decoding in favor of other codeword $\mathbf{c}_i$ whose Hamming weight is $h$, given the value of $z_1$. By applying the union bound on the RHS of (8), we get

$$\Pr(E(z_1), \mathbf{y} \in C_n(\theta) \mid z_1) \le \sum_{h=1}^{n} A_h \Pr(E_h(z_1), \mathbf{y} \in C_n(\theta) \mid z_1). \tag{9}$$

Combining (7) and (9) gives

$$\Pr\left(E(z_1) \mid z_1\right) \le \sum_h \Big\{ A_h \Pr\left(E_h(z_1), \mathbf{y} \in C_n(\theta) \mid z_1\right) \Big\} + \Pr\left(\mathbf{y} \notin C_n(\theta) \mid z_1\right). \tag{10}$$

The second term in the RHS of (10) is evaluated from (5)

$$\begin{aligned}\Pr(\mathbf{y} \notin C_n(\theta) \mid z_1) &= \Pr\left(Y \ge r_{z_1}^2 \mid z_1\right) \\ &= \int_{r_{z_1}^2}^{\infty} f_Y(y) dy \\ &= \int_{r_{z_1}^2}^{\infty} \frac{y^{\frac{n-2}{2}} e^{-\frac{y}{2\sigma^2}} U(y)}{2^{\frac{n-1}{2}} \sigma^{n-1} \Gamma\left(\frac{n-1}{2}\right)} dy.\end{aligned} \tag{11}$$

This integral can be expressed in terms of the incomplete Gamma function

$$\gamma(a, x) \triangleq \frac{1}{\Gamma(a)} \int_0^x t^{x-1} e^{-t} dt, \quad a > 0,\ x \ge 0 \tag{12}$$

and it is transformed to

$$\Pr(\mathbf{y} \notin C_n(\theta) \mid z_1) = 1 - \gamma\left(\frac{n-1}{2}, \frac{r_{z_1}^2}{2\sigma^2}\right). \tag{13}$$

Let $z_2$ designate the tangential component of the noise vector $\mathbf{z}$, which is on the plane that contains the signals $\mathbf{s}_0$, $\mathbf{s}_i$ and the origin of the space, and orthogonal to $z_1$ (see Fig. 1). Referring to the first term in the RHS of (10), it follows from the geometry in Fig. 1 that if $z_1 \le \sqrt{nE_\mathrm{s}}$ then

$$\begin{aligned}\Pr(E_h(z_1), \mathbf{y} \in C_n(\theta) \mid z_1) &= \Pr(E_h(z_1), Y \le r_{z_1}^2 \mid z_1) \\ &= \Pr\left(\beta_h(z_1) \le z_2 \le r_{z_1}, Y \le r_{z_1}^2 \mid z_1\right).\end{aligned} \tag{14}$$

Let $V \triangleq \sum_{i=3}^{n} z_i^2$, then $V = Y - z_2^2$. If $z_1 \le \sqrt{nE_\mathrm{s}}$, then we obtain the equality

$$\Pr\left(E_h(z_1), \mathbf{y} \in C_n(\theta) \mid z_1\right) = \Pr\left(\beta_h(z_1) \le z_2 \le r_{z_1}, V \le r_{z_1}^2 - z_2^2 \mid z_1\right). \tag{15}$$

The random variable $V$ is $\chi^2$ distributed with $n-2$ degrees of freedom, so its *pdf* is

$$f_V(v) = \frac{y^{\frac{n-4}{2}} e^{-\frac{y}{2\sigma^2}} U(y)}{2^{\frac{n-2}{2}} \sigma^{n-2} \Gamma\left(\frac{n-2}{2}\right)}, \quad v \ge 0 \tag{16}$$

and since the random variables $V$ and $Z_2$ are statistically independent, then if $z_1 \leq \sqrt{nE_s}$

$$\Pr\left(E_h(z_1), \mathbf{y} \in C_n(\theta) \mid z_1\right) = \int_{\beta_h(z_1)}^{r_{z_1}} \frac{e^{-\frac{z_2^2}{2\sigma^2}}}{\sqrt{2\pi}\sigma} \int_0^{r_{z_1}^2 - z_2^2} f_V(v) dv \, dz_2. \tag{17}$$

In order to obtain an upper bound on the decoding error probability, $\Pr(E)$, one should apply the statistical expectation operator on the RHS of (10) w.r.t. the radial noise component $z_1$. Referring to the upper half azimuthal cone depicted in Fig. 1 which corresponds to the case where the radial noise component satisfies the condition $z_1 \leq \sqrt{nE_s}$, the inequality $\beta_h(z_1) < r_{z_1}$ holds for the values of $h$ for which $\frac{\delta_h}{2} < \alpha_h$ where

$$\alpha_h \triangleq r\sqrt{1 - \frac{\delta_h^2}{4nE_s}}. \tag{18}$$

On the other hand, if $z_1 > \sqrt{nE_s}$, the range of integration for the component noise $z_2$ is $\beta_h(z_1) \leq z_2 \leq -r_{z_1}$ which is satisfied for all values of $h$ (since for $z_1 > \sqrt{nE_s}$, we get from (4) that $r_{z_1} < 0$ and $\beta_h(z_1) < 0$, so the inequality $\beta_h(z_1) \leq -r_{z_1}$ holds in this case for all values of $h$). Since $Z_1 \sim N(0, \sigma^2)$ where $\sigma^2 = \frac{N_0}{2}$, then the probability that the Gaussian random variable $Z_1$ exceeds $\sqrt{nE_s}$ is equal to

$$Q\left(\frac{\sqrt{nE_s}}{\sigma}\right) = Q\left(\sqrt{\frac{2nRE_b}{N_0}}\right).$$

This results in the following upper bound on the decoding error probability under ML decoding

$$\Pr(E) \leq \int_{-\infty}^{+\sqrt{nE_s}} \frac{e^{-\frac{z_2^2}{2\sigma^2}}}{\sqrt{2\pi}\sigma} \Bigg\{ \sum_{h: \frac{\delta_h}{2} < \alpha_h} \left\{ A_h \int_{\beta_h(z_1)}^{r_{z_1}} \frac{e^{-\frac{z_2^2}{2\sigma^2}}}{\sqrt{2\pi}\sigma} \int_0^{r_{z_1}^2 - z_2^2} f_V(v) dv \, dz_2 \right\} + 1 - \gamma\left(\frac{n-1}{2}, \frac{r_{z_1}^2}{2\sigma^2}\right) \Bigg\} dz_1 + Q\left(\sqrt{\frac{2nRE_b}{N_0}}\right). \tag{19}$$

The upper bound (19) is valid for all positive values of $r$. Hence, in order to achieve the tightest upper bound of the form (19) one should set to zero the partial derivative of the RHS of (19) w.r.t. $r_{z_1}$. After straightforward algebra the following optimization equation for the optimal value of $r$ is obtained [9]:

$$\begin{cases} \displaystyle\sum_{h: \frac{\delta_h}{2} < \alpha_h} A_h \int_0^{\theta_h} \sin^{n-3}\phi \, d\phi = \frac{\sqrt{\pi}\,\Gamma(\frac{n-2}{2})}{\Gamma(\frac{n-1}{2})} \\ \theta_h = \cos^{-1}\left(\frac{\delta_h}{2\alpha_h}\right) \end{cases} \tag{20}$$

where $\alpha_h$ is given in (18). A proof for the existence and uniqueness of a solution $r$ to the optimization equation (20) was provided in [11, Appendix B], together with an efficient algorithm to solve this equation numerically. In order to derive an upper bound on the bit error probability, let $A_{w,h}$ designate the corresponding coefficient in the IOWEF which is the number of codewords which are encoded by information bits whose number of ones is equal to $w$ (where $0 \leq w \leq nR$) and whose Hamming weights (after encoding) are equal to $h$, and define

$$A'_h \triangleq \sum_{w=1}^{nR} \left(\frac{w}{nR}\right) A_{w,h}, \quad h = 0, \ldots, n. \tag{21}$$

In [11, Appendix C], Sason and Shamai derive an upper bound on the bit error probability by replacing the distance spectrum $\{A_h\}$ in (19) and (20) with the sequence $\{A'_h\}$, and show some properties on the resulting bound on the bit error probability.

## 2.3 Improved Tangential-Sphere Bound

In [17], Yousefi and Mehrabian derive a new upper bound on the block error probability of binary linear block codes whose transmission takes place over a binary-input AWGN channel, and which are coherently detected and ML decoded. This upper bound, which is called improved tangential-sphere bound (ITSB) is based on inequality (3), where the region $\mathcal{R}$ is the same as of the TSB (i.e., an $n$-dimensional circular cone). To this end, the ITSB is obtained by applying a Bonferroni-type inequality of the second order [5, 8] (instead of the union bound) to get an upper bound on the joint probability of decoding error and the event that the received vector falls within the corresponding conical region around the transmitted signal vector.

The basic idea in [17] relies on Hunter's bound which states that if $\{E_i\}_{i=1}^M$ designates a set of $M$ events, and $E_i^{\rm c}$ designates the complementary event of $E_i$, then

$$\begin{aligned}\Pr\left(\bigcup_{i=1}^{M} E_i\right) &= \Pr(E_1) + \Pr(E_2 \cap E_1^{\rm c}) + \ldots + \Pr(E_M \cap E_{M-1}^{\rm c} \ldots \cap E_1^{\rm c}) \\ &\leq \Pr(E_1) + \sum_{i=2}^{M} \Pr(E_i \cap E_{\hat{i}}^{\rm c}). \end{aligned} \tag{22}$$

where the indices $\hat{i} \in \{1, 2, \ldots i-1\}$ are chosen arbitrarily for $i \in \{2, \ldots, M\}$. Clearly, the upper bound (22) is tighter than the union bound. The LHS of (22) is invariant to the ordering of the events (since it only depends on the union of these events), while the RHS of (22) depends on this ordering. Hence, the tightest bound of the form (22) is obtained by choosing the optimal indices ordering $i \in \{1, 2, \ldots, M\}$ and $\hat{i} \in \{1, 2, \ldots, i-1\}$. Let us designate by $\Pi(1, 2, \ldots, M) = \{\pi_1, \pi_2, \ldots, \pi_M\}$ an arbitrary permutation among the $M!$ possible permutations of the set $\{1, 2, \ldots, M\}$ (i.e., a permutation of the indices of the events $E_1$ to $E_M$), and let $\Lambda = (\lambda_2, \lambda_3, \ldots \lambda_M)$ designate an arbitrary sequence of integers where $\lambda_i \in \{\pi_1, \pi_2, \ldots \pi_{i-1}\}$. Then, the tightest form of of the bound in (22) is given by

$$\Pr\left(\bigcup_{i=1}^{M} E_i\right) \leq \min_{\Pi,\Lambda} \left\{ \Pr(E_{\pi_1}) + \sum_{i=2}^{M} \Pr(E_{\pi_i} \cap E_{\lambda_i}^{\rm c}) \right\}. \tag{23}$$

Similar to the TSB, the derivation of the ITSB originates from the upper bound (7) on the conditional decoding error probability, given the radial component ($z_1$) of the noise vector (see Fig. 1). In [17], it is proposed to apply the upper bound (23) on the RHS of (8) which for an arbitrary permutation $\{\pi_1, \pi_2, \ldots, \pi_M\}$ and a corresponding sequence of integers $(\lambda_2, \lambda_3, \ldots \lambda_{M-1})$ as above, gives

$$\begin{aligned}\Pr\left(\bigcup_{i=1}^{M-1} E_{0\to i}, \mathbf{y} \in C_n(\theta) \mid z_1\right) \leq \min_{\Pi,\Lambda} \Bigg\{ &\Pr(E_{0\to\pi_1}, \mathbf{y} \in C_n(\theta) \mid z_1) \\ &+ \sum_{i=2}^{M-1} \Pr(E_{0\to\pi_i}, E_{0\to\lambda_i}^{\rm c}, \mathbf{y} \in C_n(\theta) \mid z_1) \Bigg\} \end{aligned} \tag{24}$$

where $E_{0\to j}$ designates the pairwise error event where the decoder decides on codeword $\mathbf{c}_j$ rather than the transmitted codeword $\mathbf{c}_0$. As indicated in [15, 17], the optimization problem of (24) is prohibitively complex. In order to simplify it, Yousefi and Mehrabian suggest to choose $\pi_1 = \lambda_i = i_{\min}$ for all $i = 2, \ldots, M-1$, where $i_{\min}$ designates the index of a codeword which is closest (in terms of Euclidian distance) to the transmitted signal vector $\mathbf{s}_0$. Since the code is linear and the channel is memoryless and symmetric, one can assume without any loss of generality that the all-zero codeword is transmitted. Moreover, since we deal with antipodal modulation, then $w_{\mathrm{H}}(\mathbf{c}_{i_{\min}}) = d_{\min}$ where $d_{\min}$ is the minimum distance of the code. Hence, by this specific choice of $\pi_1$ and $\Lambda$ (which in general loosen the tightness of the bound in (24)), the ordering of the indices $\{\pi_2, \ldots, \pi_M\}$ is irrelevant, and one can omit the optimization over $\Pi$ and $\Lambda$. The above simplification results in the following inequality:

$$\begin{aligned}\Pr(E|z_1) \leq & \Pr\left(E_{0\to i_{\min}}, \mathbf{y} \in C_n(\theta) \mid z_1\right) \\ & + \sum_{i=2}^{M-1} \Pr(E_{0\to i}, E^{\mathrm{c}}_{0\to i_{\min}}, \mathbf{y} \in C_n(\theta) \mid z_1) + \Pr\left(\mathbf{y} \notin C_n(\theta) \mid z_1\right). \end{aligned} \tag{25}$$

Based on Fig. 1, the first and the third terms in the RHS of (25) can be evaluated in similarity with the TSB, and we get

$$\Pr\left(E_{0\to i_{\min}}, \mathbf{y} \in C_n(\theta) \mid z_1\right) = \Pr(\beta_{\min}(z_1) \leq z_2 \leq r_{z_1},\ V < r_{z_1}^2 - z_2^2 \mid z_1) \tag{26}$$

$$\Pr(\mathbf{y} \notin C_n(\theta) \mid z_1) = 1 - \gamma\left(\frac{n-1}{2}, \frac{r_{z_1}^2}{2\sigma^2}\right) \tag{27}$$

where

$$\beta_{\min}(z_1) = \left(\sqrt{nE_{\mathrm{s}}} - z_1\right)\sqrt{\frac{d_{\min}}{n - d_{\min}}}, \tag{28}$$

$z_2$ is the tangential component of the noise vector $\mathbf{z}$, which is on the plane that contains the signals $\mathbf{s}_0$, $\mathbf{s}_{i_{\min}}$ and the origin (see Fig. 1), and the other parameters are introduced in (4).

For expressing the probabilities of the form $\Pr(E_{0\to i}, E^{\mathrm{c}}_{0\to i_{\min}}, \mathbf{y} \in C_n(\theta) \mid z_1)$ encountered in the RHS of (25), we use the three-dimensional geometry in Fig. 2-(a). The BPSK modulated signals $\mathbf{s}_0$, $\mathbf{s}_i$ and $\mathbf{s}_j$ are all on the surface of a hyper-sphere centered at the origin and with radius $\sqrt{nE_{\mathrm{s}}}$. The planes $P_1$ and $P_2$ are constructed by the points $(\mathbf{o}, \mathbf{s}_0, \mathbf{s}_i)$ and $(\mathbf{o}, \mathbf{s}_0, \mathbf{s}_j)$, respectively. In the derivation of the ITSB, Yousefi and Mehrabian choose $\mathbf{s}_j$ to correspond to codeword $\mathbf{c}_j$ with Hamming weight $d_{\min}$. Let $z_3'$ be the noise component which is orthogonal to $z_1$ and which lies on the plane $P_2$ (see Fig 2-a). Based on the geometry in Fig. 2-a (the probability of the event $E^c_{0\to j}$ is the probability that $\mathbf{y}$ falls in the dashed area) we obtain the following equality if $z_1 \leq \sqrt{nE_{\mathrm{s}}}$:

$$\begin{aligned}&\Pr(E_{0\to i}, E^c_{0\to i_{\min}}, \mathbf{y} \in C_n(\theta) \mid z_1) \\ &= \Pr\left(\beta_i(z_1) \leq z_2 \leq r_{z_1},\ -r_{z_1} \leq z_3' \leq \beta_{\min}(z_1),\ Y < r_{z_1}^2 \mid z_1\right).\end{aligned} \tag{29}$$

Furthermore, from the geometry in Fig. 2-b, it follows that

$$z_3' = z_3 \sin\phi + z_2 \cos\phi. \tag{30}$$

where $z_3$ is the noise component which is orthogonal to both $z_1$ and $z_2$, and which resides in the three-dimensional space that contains the signal vectors $\mathbf{s}_0$, $\mathbf{s}_i$, $\mathbf{s}_{i_{\min}}$ and the origin. Plugging (30) into the condition $-r_{z_1} \leq z_3' \leq \beta_{\min}(z_1)$ in (29) yields the condition $-r_{z_1} \leq z_3 \leq \min\{l(z_1,z_2), r_{z_1}\}$ where

$$l(z_1,z_2) = \frac{\beta_{\min}(z_1) - \rho z_2}{\sqrt{1-\rho^2}} \tag{31}$$

and $\rho = \cos\phi$ is the correlation coefficient between planes $P_1$ and $P_2$. Let $W = \sum_{i=4}^{n} z_i^2$, then if $z_1 \leq \sqrt{nE_{\text{s}}}$

$$\begin{aligned}&\Pr(E_{0\to i}, E^c_{0\to i_{\min}}, \mathbf{y} \in C_n(\theta) \mid z_1) \\ &= \Pr\left(\beta_i(z_1) \leq z_2 \leq r_{z_1},\ -r_{z_1} \leq z_3 \leq \min\{l(z_1,z_2), r_{z_1}\},\ W < r_{z_1}^2 - z_2^2 - z_3^2 \mid z_1\right).\end{aligned} \tag{32}$$

The random variable $W$ is Chi-squared distributed with $n-3$ degrees of freedom, so its *pdf* is given by

$$f_W(w) = \frac{w^{\frac{n-5}{2}} e^{-\frac{w}{2\sigma^2}} U(w)}{2^{\frac{n-3}{2}} \sigma^{n-3} \Gamma\left(\frac{n-3}{2}\right)}, \quad w \geq 0. \tag{33}$$

Since the probabilities of the form $\Pr(E_{0\to i}, E^c_{0\to i_{\min}}, \mathbf{y} \in C_n(\theta) \mid z_1)$ depend on the correlation coefficients between the planes $(\mathbf{o}, \mathbf{s}_0, \mathbf{s}_{i_{\min}})$ and $(\mathbf{o}, \mathbf{s}_0, \mathbf{s}_i)$, the overall upper bound requires the characterization of the global geometrical properties of the code and not only the distance spectrum. To circumvent this problem and obtain an upper bound which is solely depends on the distance spectrum of the code, it is suggested in [17] to loosen the bound as follows. It is shown [16, Appendix B] that the correlation coefficient $\rho$, corresponding to codewords with Hamming weights $d_i$ and $d_j$ satisfies

$$-\min\left\{\sqrt{\frac{d_i d_j}{(n-d_i)(n-d_j)}}, \sqrt{\frac{(n-d_i)(n-d_j)}{d_i d_j}}\right\} \leq \rho \leq \frac{\min(d_i,d_j)[n-\max(d_i,d_j)]}{\sqrt{d_i d_j (n-d_i)(n-d_j)}}. \tag{34}$$

Moreover, the RHS of (32) is shown to be a monotonic decreasing function of $\rho$ (see [17, Appendix 1]). Hence, one can omit the dependency in the geometry of the code (and loosen the upper bound) by replacing the correlation coefficients in (32) with their lower bounds which solely depend on the weights of the codewords. In the derivation of the ITSB, we consider the correlation coefficients between two planes which correspond to codewords with Hamming weights $d_i = h,\ h \geq n$ and $d_j = d_{\min}$. Let

$$\begin{aligned}\rho_h &\triangleq -\min\left\{\sqrt{\frac{h d_{\min}}{(n-h)(n-d_{\min})}}, \sqrt{\frac{(n-h)(n-d_{\min})}{h d_{\min}}}\right\} \\ &= -\sqrt{\frac{h d_{\min}}{(n-h)(n-d_{\min})}},\end{aligned} \tag{35}$$

where the last equality follows directly from the basic property of $d_{\min}$ as the minimum distance of the code. From (25)–(26) and averaging w.r.t. $Z_1$, one gets the following upper bound on the decoding error probability:

$$\begin{aligned}
\Pr(E) \le \Pr\left(z_1 \le \sqrt{nE_\mathrm{s}},\ \beta_{\min}(z_1) \le z_2 \le r_{z_1}, V \le r_{z_1}^2 - z_2^2\right)& \\
+ \sum_{h=d_{\min}}^{n} A_h \Pr\Big(z_1 \le \sqrt{nE_\mathrm{s}},\ \beta_h(z_1) \le z_2 \le r_{z_1},& \\
- r_{z_1} \le z_3 \le \min\{l_h(z_1,z_2), r_{z_1}\}, W \le r_{z_1}^2 - z_2^2 - z_3^2\Big)& \\
+ \Pr\left(z_1 \le \sqrt{nE_\mathrm{s}},\ Y \ge r_{z_1}^2\right) + \Pr(z_1 > \sqrt{nE_\mathrm{s}})& \qquad (36)
\end{aligned}$$

where the parameter $l_h(z_1, z_2)$ is simply $l(z_1, z_2)$ in (31) with $\rho$ replaced by $\rho_h$, i.e.,

$$l_h(z_1, z_2) \triangleq \frac{\beta_{\min}(z_1) - \rho_h z_2}{\sqrt{1-\rho_h^2}}. \qquad (37)$$

Using the probability density functions of the random variables in the RHS of (36), and since the random variables $Z_1, Z_2, Z_3$ and $W$ are statistically independent, the final form of the ITSB is given by

$$\begin{aligned}
P_\mathrm{e} \le \int_{-\infty}^{\sqrt{nE_\mathrm{s}}} \Bigg[\int_{\beta_{\min}}^{r_{z_1}} f_{Z_2}(z_2) \int_0^{r_{z_1}^2 - z_2^2} f_V(v)dv \cdot dz_2& \\
+ \sum_{h:\beta_h(z_1) < r_{z_1}} \left(A_h \int_{\beta_h(z_1)}^{r_{z_1}} \int_{-r_{z_1}}^{\min\{l_h(z_1,z_2), r_{z_1}\}} f_{Z_2,Z_3}(z_2,z_3) \int_0^{r_{z_1}^2 - z_2^2 - z_3^2} f_W(w)dw \cdot dz_2 \cdot dz_3\right)& \\
+ 1 - \gamma\left(\frac{n-1}{2}, \frac{r_{z_1}^2}{2\sigma^2}\right)\Bigg] f_{Z_1}(z_1)dz_1 + Q\left(\sqrt{\frac{2nRE_\mathrm{b}}{N_0}}\right).& \qquad (38)
\end{aligned}$$

Note that $V \triangleq \sum_{i=3}^n z_i^2$ and $W \triangleq \sum_{i=4}^n z_i^2$ are Chi-squared distributed with $(n-2)$ and $(n-3)$ degrees of freedom, respectively.

## 2.4 Added-Hyper-Plane (AHP) Bound

In [16], Yousefi and Khandani introduce a new upper bound on the ML decoding block error probability, called the added hyper plane (AHP) bound. In similarity with the ITSB, the AHP bound is based on using the Hunter bound (22) as an upper bound on the LHS of (9), which results in the inequality (24). The complex optimization problem in (24), however, is treated differently. Let us denote by $\mathcal{I}_w$ the set of the indices of the codewords of $\mathcal{C}$ with Hamming weight $w$. For $i \in \{1, 2, \ldots, M\} \setminus \mathcal{I}_w$, let $\{j_i\}$ be a sequence of integers chosen from the set $\mathcal{I}_w$. Then the following upper bound holds

$$\begin{aligned}
&\Pr\left(E(z_1), \mathbf{y} \in C_n(\theta) \mid z_1\right) \\
&\le \min_{w, \mathcal{J}_w} \left\{ \Pr\left(\bigcup_{j \in \mathcal{I}_w} \{E_{0\to j}\}, \mathbf{y} \in C_n(\theta) \mid z_1\right) + \sum_{i \in \{1,\ldots,M-1\} \setminus \mathcal{I}_w} \Pr\left(E_{0\to i}, E_{0\to j_i}^c, \mathbf{y} \in C_n(\theta) \mid z_1\right) \right\}. \qquad (39)
\end{aligned}$$

The probabilities inside the summation in the RHS of (39) are evaluated in a similar manner to the probabilities in the LHS of (29). From the analysis in Section 2.3 and the geometry in Fig. 2-(b), it is clear that the aforementioned probabilities depend on the correlation coefficients between the planes $(\mathbf{o}, \mathbf{s}_0, \mathbf{s}_i)$ and $(\mathbf{o}, \mathbf{s}_0, \mathbf{s}_{j_i})$. Hence, in order to compute the upper bound (39), one has to know the geometrical characterization of the Voronoi regions of the codewords. To obtain an upper bound requiring only the distance spectrum of the code, Yousefi and Khandani suggest to extend the codebook by adding all the $\binom{n}{w} - A_w$ $n$-tuples with Hamming weight $w$ (i.e., the extended code contains all the binary vectors of length $n$ and Hamming weight $w$). Let us designate the new code by $\mathcal{C}_w$ and denote its codewords by

$$\mathbf{c}_i^w, \quad i \in \left\{0, 1, \ldots, M + \binom{n}{w} - A_w - 1\right\}.$$

The new codebook is not necessarily linear, and all possible correlation coefficients between two codewords with Hamming weight $i$, where $i \in \{d_{\min}, \ldots d_{\max}\}$, and $w$ are available. Thus, for each layer of the codebook, one can choose the *largest available correlation*[1] $\rho$ with respect to any possible $n$-tuple binary vector of Hamming weight $w$. Now one may find the optimum layer at which the codebook extension is done, i.e., finding the optimum $w \in \{1, 2, \ldots n\}$ which yields the tightest upper bound within this form. We note that the resulting upper bound is not proved to be uniformly tighter than the TSB, due to the extension of the code. The maximum correlation coefficient between two codewords of Hamming weight $d_i$ and $d_j$ is introduced in the RHS of (34) (see [16]). Let us designate the maximal possible correlation coefficient between two $n$-tuples with Hamming weights $w$ and $h$ by $\rho_{w,h}$ , i.e.,

$$\rho_{w,h} = \frac{\min(h, w)[n - \max(h, w)]}{\sqrt{hw(n-h)(n-w)}} \ , \quad w \neq h. \tag{40}$$

By using the same bounding technique of the ITSB, and replacing the correlation coefficients with their respective upper bounds, $\rho_{w,h}$, (39) gets the form

$$\begin{aligned} \Pr\left(E(z_1), \mathbf{y} \in C_n(\theta) \mid z_1\right) \leq \min_w \Bigg\{ & \Pr\Bigg( \bigcup_{j: w_{\mathrm{H}}(\mathbf{c}_j^w) = w} \{E_{0 \to j}\}, \mathbf{y} \in C_n(\theta) \mid z_1 \Bigg) \\ & + \sum_{h \neq w} A_h \Pr\left(Y \leq r_{z_1}^2, \beta_h(z_1) \leq z_2,\ z_3 \leq l_{w,h}(z_1, z_2) \mid z_1\right) \Bigg\} \end{aligned} \tag{41}$$

where

$$l_{w,h}(z_1, z_2) = \frac{\beta_w(z_1) - \rho_{w,h} z_2}{\sqrt{1 - \rho_{w,h}^2}}. \tag{42}$$

[1]The RHS of (39) is a monotonically decreasing function of $\rho$, as noted in [17].

Now, applying Hunter bound on the first term in the RHS of (41) yields

$$\Pr\left(\bigcup_{j:w_{\mathrm{H}}(\mathbf{c}_j^w)=w} E_{0\to j},\ \mathbf{y}\in C_n(\theta)\mid z_1\right)$$
$$\leq \Pr(E_{0\to l_0},\mathbf{y}\in C_n(\theta)\mid z_1)+\sum_{i=1}^{\binom{n}{w}-1}\Pr(E_{0\to l_i},E^c_{0\to \hat{l}_i}\mathbf{y}\in C_n(\theta)\mid z_1) \tag{43}$$

where $\{l_i\},\ i\in\left\{0,1,\ldots,\binom{n}{w}-1\right\}$ is a sequence which designates the indices of the codewords of $\mathcal{C}_w$ with Hamming weight $w$ with an arbitrary order, and $\hat{l}_i\in(l_0,l_1,\ldots,l_{i-1})$. In order to obtain the tightest bound on the LHS of (43) in this approach, one has to order the error events such that the correlation coefficients which correspond to codewords $\mathbf{c}_{l_i}$ and $\mathbf{c}_{\hat{l}_i}$ get their maximum available value, which is $1-\frac{n}{w(n-w)}$ [16, Appendix D]. Let us designate this value by $\rho_{w,w}$ ,i.e.,

$$\rho_{w,w}=1-\frac{n}{w(n-w)}\quad ,w\notin\{0,n\}.$$

Hence, based on the geometry in Fig. 2, if $z_1\leq\sqrt{nE_{\mathrm{s}}}$, we can rewrite (43) as

$$\Pr\left(\bigcup_{j:w_{\mathrm{H}}(\mathbf{c}_j^w)=w} E_{0\to j},\mathbf{y}\in C_n(\theta)\mid z_1\right)$$
$$\leq \Pr\left(\beta_w(z_1)\leq z_2\leq r_{z_1},V\leq r_{z_1}^2-z_2^2\mid z_1\right)$$
$$+\left[\binom{n}{w}-1\right]\Pr\left(\beta_w(z_1)\leq z_2\leq r_{z_1},-r_{z_1}\leq z_3\leq\min\{l_{w,w}(z_1,z_2),r_{z_1}\},W\leq r_{z_1}^2-z_2^2-z_3^2\mid z_1\right) \tag{44}$$

where

$$l_{w,w}(z_1,z_2)=\frac{\beta_w(z_1)-\rho_{w,w}z_2}{\sqrt{1-\rho_{w,w}^2}}. \tag{45}$$

By replacing the first term in the RHS of (41) with the RHS of (44), plugging the result in (7) and averaging w.r.t. $Z_1$ finally gives the following upper bound on the block error probability:

$$\Pr(E)\leq\min_w\Bigg\{\Pr\left(z_1\leq\sqrt{nE_{\mathrm{s}}},\ \beta_w(z_1)\leq z_2\leq r_{z_1},V\leq r_{z_1}^2-z_2^2\right)$$
$$+\binom{n}{w}\Pr\left(z_1\leq\sqrt{nE_{\mathrm{s}}},\ \beta_w(z_1)\leq z_2\leq r_{z_1},-r_{z_1}\leq z_3\leq\min\{l_{w,w}(z_1,z_2),r_{z_1}\},W\leq r_{z_1}^2-z_2^2-z_3^2\right)$$
$$+\sum_{h\neq w}A_h\Pr\left(z_1\leq\sqrt{nE_{\mathrm{s}}},\ \beta_h(z_1)\leq z_2\leq r_{z_1},-r_{z_1}\leq z_3\leq\min\{l_{w,h}(z_1,z_2),r_{z_1}\},W\leq r_{z_1}^2-z_2^2-z_3^2\right)\Bigg\}$$
$$+\Pr\left(z_1\leq\sqrt{nE_{\mathrm{s}}},\ Y\geq r_{z_1}^2\right)\Bigg\}+\Pr\left(z_1>\sqrt{nE_{\mathrm{s}}}\right). \tag{46}$$

Rewriting the RHS of (46) in terms of probability density functions, the AHP bound gets the form

$$P_{\mathrm{e}} \le \min_{w} \Bigg\{ \int_{-\infty}^{\sqrt{nE_{\mathrm{s}}}} \Bigg[ \int_{\beta_w(z_1)}^{r_{z_1}} f_{Z_2}(z_2) \int_0^{r_{z_1}^2 - z_2^2} f_V(v) dv \cdot dz_2$$
$$+ \binom{n}{w} \int_{\beta_w(z_1)}^{r_{z_1}} \int_{-r_{z_1}}^{\min\{l_{w,w}(z_1,z_2), r_{z_1}\}} f_{Z_2,Z_3}(z_2,z_3) \int_0^{r_{z_1}^2 - z_2^2 - z_3^2} f_W(w) dw \cdot dz_2 \cdot dz_3$$
$$+ \sum_{\substack{h\,:\,\beta_h(z_1) < r_{z_1} \\ h \neq w}} \left( A_h \int_{\beta_h(z_1)}^{r_{z_1}} \int_{-r_{z_1}}^{\min\{l_{w,h}(z_1,z_2), r_{z_1}\}} f_{Z_2,Z_3}(z_2,z_3) \int_0^{r_{z_1}^2 - z_2^2 - z_3^2} f_W(w) dw \cdot dz_2 \cdot dz_3 \right)$$
$$+ 1 - \gamma\left(\frac{n-1}{2}, \frac{r_{z_1}^2}{2\sigma^2}\right) \Bigg] f_{Z_1}(z_1) dz_1 \Bigg\} + Q\left(\sqrt{\frac{2nRE_{\mathrm{b}}}{N_0}}\right) \qquad (47)$$

where $V$ and $W$ are introduced at the end of Section 2.3 (after Eq. (38)), and the last term in (47) follows from (13).

# 3 The Error Exponents of the ITSB and AHP Bounds

The ITSB and the AHP bound were originally derived in [16, 17] as upper bounds on the ML decoding error probability of *specific* binary linear block codes. In the following, we discuss the tightness of the new upper bounds for ensemble of codes, as compared to the TSB. The following lemma is also noted in [17].

**Lemma 1.** Let $\mathcal{C}$ be a binary linear block code, and let us denote by ITSB($\mathcal{C}$) and TSB($\mathcal{C}$) the ITSB and TSB, respectively, on the decoding error probability of $\mathcal{C}$. Then

$$\mathrm{ITSB}(\mathcal{C}) \le \mathrm{TSB}(\mathcal{C}).$$

*Proof.* Since $\Pr(A, B) \le \Pr(A)$ for arbitrary events $A$ and $B$, the lemma follows immediately by comparing the bounds in the RHS of (10) and (25), reffering to the TSB and the ITSB, respectively. □

**Corollary 1.** The ITSB can not exceed the value of the TSB referring to the average error probability of an arbitrary ensemble of binary linear block codes.

**Lemma 2.** The AHP bound is asymptotically (as we let the block length tend to infinity) at least as tight as the TSB.

*Proof.* To show this, we refer to (46), where we choose the layer $w$ at which the extension of the code is done to be $n$. Hence, the extended code contains at most one codeword with Hamming weight $n$ more than the original code, which has no impact on the error probability for infinitely long codes. The resulting upper bound is evidently not tighter than the AHP (which carries an optimization over $w$), and it is at least as tight as the TSB (since the joint probability of two events cannot exceed the probabilities of these individual events). □

The extension of Lemma 2 to ensembles of codes is straightforward (by taking the expectation over the codes in an ensemble, the same conclusion in Lemma 2 holds also for ensembles). From the above, it is evident that the error exponents of both the AHP bound and the ITSB cannot be below the error exponent of the TSB. In the following, we introduce a lower bound on both the ITSB and the AHP bound. It serves as an intermediate stage to get our main result.

**Lemma 3.** Let $\mathcal{C}$ designate an ensemble of linear codes of length $n$, whose transmission takes place over an AWGN channel. Let $A_h$ be the number of codewords of Hamming weight $h$, and let $\mathbb{E}_{\mathcal{C}}$ designate the statistical expectation over the codebooks of an ensemble $\mathcal{C}$. Then both the ITSB and AHP upper bounds on the average ML decoding error probability of $\mathcal{C}$ are lower bounded by the function $\psi(\mathcal{C})$ where

$$\begin{aligned}\psi(\mathcal{C}) \triangleq \min_{w} \Bigg\{ \mathbb{E}_{\mathcal{C}} \Bigg[ &\Pr\left(z_1 \leq \sqrt{nE_{\mathrm{s}}},\ \beta_w(z_1) \leq z_2 \leq r_{z_1}, V \leq r_{z_1}^2 - z_2^2\right) \\ &+ \sum_h \Bigg\{ A_h \Pr\Big(z_1 \leq \sqrt{nE_{\mathrm{s}}},\ \beta_h(z_1) \leq z_2 \leq r_{z_1}, \\ &\qquad - r_{z_1} \leq z_3 \leq \min\{l_{w,h}(z_1, z_2), r_{z_1}\}, W \leq r_{z_1}^2 - z_2^2 - z_3^2\Big) \Bigg\} \\ &+ \Pr\left(z_1 \leq \sqrt{nE_{\mathrm{s}}},\ Y \geq r_{z_1}^2\right) \Bigg] \Bigg\} \end{aligned} \tag{48}$$

and $l_{w,h}(z_1, z_2)$ is defined in (42).

*Proof.* By comparing (46) with (48), it is easily verified that the RHS of (48) is not larger than the RHS of (46) (actually, the RHS of (48) is just the AHP *without* any extension of the code). Referring to the ITSB, we get

$$\begin{aligned}\mathrm{ITSB}(\mathcal{C}) = \mathbb{E}_{\mathcal{C}} \Bigg[ &\Pr\left(z_1 \leq \sqrt{nE_{\mathrm{s}}},\ \beta_{\min}(z_1) \leq z_2 \leq r_{z_1}, V \leq r_{z_1}^2 - z_2^2\right) \\ &+ \sum_h \Bigg\{ A_h \Pr\Big(z_1 \leq \sqrt{nE_{\mathrm{s}}},\ \beta_h(z_1) \leq z_2 \leq r_{z_1}, \\ &\qquad - r_{z_1} \leq z_3 \leq \min\{l_h(z_1, z_2), r_{z_1}\},\ W \leq r_{z_1}^2 - z_2^2 - z_3^2\Big) \Bigg\} \\ &+ \Pr\left(z_1 \leq \sqrt{nE_{\mathrm{s}}},\ Y \geq r_{z_1}^2\right) \Bigg] + \Pr\left(z_1 > \sqrt{nE_{\mathrm{s}}}\right) \\ \geq \min_{w} \Bigg\{ \mathbb{E}_{\mathcal{C}} \Bigg[ &\Pr\left(z_1 \leq \sqrt{nE_{\mathrm{s}}},\ \beta_w(z_1) \leq z_2 \leq r_{z_1}, V \leq r_{z_1}^2 - z_2^2\right) \\ &+ \sum_h \Bigg\{ A_h \Pr\Big(z_1 \leq \sqrt{nE_{\mathrm{s}}},\ \beta_h(z_1) \leq z_2 \leq r_{z_1}, \\ &\qquad - r_{z_1} \leq z_3 \leq \min\{l_{w,h}(z_1, z_2), r_{z_1}\}, W \leq r_{z_1}^2 - z_2^2 - z_3^2\Big) \Bigg\} \\ &+ \Pr\left(z_1 \leq \sqrt{nE_{\mathrm{s}}},\ Y \geq r_{z_1}^2\right) \Bigg] \Bigg\} + \Pr\left(z_1 > \sqrt{nE_{\mathrm{s}}}\right) \\ > \psi(\mathcal{C}). \qquad & \end{aligned} \tag{49}$$

The first inequality holds since the ITSB is a monotonically decreasing function w.r.t. the correlation coefficients (see Appendix A.3). The equality in (49) is due to the linearity of the function in (49) w.r.t. the distance spectrum, on which the expectation operator is applied, and the last transition follows directly from (48). □

In [16] and [17], the RHS of (46) and (36), respectively, were evaluated by integrals, which results in the upper bounds (47) and (38). In [1, Section D], Divsalar introduced an alternative way to obtain a simple, yet asymptotically identical, version of the TSB by using the Chernoff bounding technique. Using this technique we obtain the exponential version of $\psi(\mathcal{C})$. In the following, We use the following notation [1]:

$$c \triangleq \frac{E_{\text{s}}}{N_0}, \quad \delta \triangleq \frac{h}{n}, \quad \Delta \triangleq \sqrt{\frac{\delta}{1-\delta}}, \quad r(\delta) \triangleq \frac{\ln(A_h)}{n}$$

where for the sake of clear writing we denote the average spectrum of the ensemble by $A_h$. We now state the main result of this paper.

**Theorem 1. (The error exponent of the AHP and the ITSB bounds coincide with the error exponent of the TSB)** The upper bounds ITSB, AHP and the TSB have the same error exponent, which is

$$E(c) = \min_{0<\delta\leq 1} \left\{ \frac{1}{2} \ln\left(1 - \gamma + \gamma e^{-2r(\delta)}\right) + \frac{\gamma \Delta^2 c}{1+\gamma\Delta^2} \right\} \tag{50}$$

where

$$\gamma = \gamma(\delta) \triangleq \frac{1-\delta}{\delta} \left[ \sqrt{\frac{c}{c_0(\delta)} + (1+c)^2 - 1} - (1+c) \right] \tag{51}$$

and

$$c_0(\delta) \triangleq \left(1 - e^{-2r(\delta)}\right) \frac{1-\delta}{2\delta}. \tag{52}$$

*Proof.* The exponential version of $\psi(\mathcal{C})$ in (48) is identical to the exponential version of the TSB (see Appendices A.1 and A.2). Since $\psi(\mathcal{C})$ does not exceed the AHP and the ITSB, this implies that the error exponents of the AHP and the ITSB are not larger than the error exponent of the TSB. On the other hand, from Lemmas 1 and 2 it follows that asymptotically, both the AHP and the ITSB are at least as tight as the TSB, so their error exponents are at least as large as the error exponent of the TSB. Combining these results we obtain that the error exponent of the ITSB, AHP and the TSB are all identical. In [1], Divsalar shows that the error exponent of the TSB is determined by (50)–(52), which concludes the proof of the theorem. □

**Remark 1.** The bound on the bit error probability in [11] is exactly the same as the TSB on the block error probability by Poltyrev [9], except that the average distance spectrum $\{A_h\}$ of the ensemble is now replaced by the sequence $\{A'_h\}$ where

$$A'_h = \sum_{w=0}^{nR} \left(\frac{w}{nR}\right) A_{w,h}\,, \quad h \in \{0, \ldots, n\}$$

and $A_{w,h}$ denotes the average number of codewords encoded by information bits of Hamming weight $w$ and having a Hamming weight (after encoding) which is equal to $h$. Since $A_h = \sum_{w=0}^{nR} A_{w,h}$, then

$$\frac{A_h}{nR} \leq A'_h \leq A_h \ , \quad h \in \{0, \ldots, n\}.$$

The last inequality therefore implies that the replacement of the distance spectrum $\{A_h\}$ by $\{A'_h\}$ (for the analysis of the bit error probability) does not affect the asymptotic growth rate of $r(\delta)$ where $\delta \triangleq \frac{h}{n}$, and hence, the error exponents of the TSB on the block and bit error probabilities coincide.

**Remark 2.** In [19], Zangl and Herzog suggest a modification of the TSB on the bit error probability. Their basic idea is tightening the bound on the bit error probability when the received vector $\mathbf{y}$ falls outside the cone $\mathcal{R}$ in the RHS of (3) (see Fig. 1). In the derivation of the version of the TSB on the bit error probability, as suggested by Sason and Shamai [11], the conditional bit error probability in this case was upper bounded by 1, where Zangl and Herzog [19] refine the bound and provide a tighter bound on the conditional bit error probability when the vector $\mathbf{y}$ falls in the bad region (i.e., when it is outside the cone in Fig. 1). Though this modification tightens the bound on the bit error probability at low SNR (as exemplified in [19] for some short linear block codes), it has no effect on the error exponent. The reason is simply because the conditional bit error probability in this case cannot be below $\frac{1}{nR}$ (i.e., one over the dimension of the code), so the bound should still possess the same error exponent. This shows that the error exponent of the TSB versions on the bit error probability, as suggested in [11] and [19], coincide.

**Corollary 2.** The error exponents of the TSB on the bit error probability coincides with the error exponent of the TSB on the block error probability. Moreover, the error exponents of the TSB on the bit error probability, as suggested by Sason and Shamai [11] and refined by Zangl and Herzog [19], coincide. The common value of these error exponents is explicitly given in Theorem 1.

# 4 Summary and Conclusions

The tangential-sphere bound (TSB) of Poltyrev [9] often happens to be the tightest upper bound on the ML decoding error probability of block codes whose transmission takes place over a binary-input AWGN channel. However, in the random coding setting, it fails to reproduce the random coding error exponent [6] while the second version of the Duman and Salehi (DS2) bound does [3, 12]. The larger the code rate is, the more significant becomes the gap between the error exponent of the TSB and the random coding error exponent of Gallager [6] (see Fig. 3, and the plots in [9, Figs. 2–4]). In this respect, we note that the expression for the error exponent of the TSB, as derived by Divsalar [1], is significantly easier for numerical calculations than the original expression of this error exponent which was provided by Poltyrev [9, Theorem 2]. Moreover, the analysis made by Divsalar is more general in the sense that it applies to an arbitrary ensemble, and not only to the ensemble of fully random block codes.

In this paper, we consider some recently introduced performance bounds which suggest an improvement over the TSB. These bounds rely solely on the distance spectrum of the code (or their input-output weight enumerators for the analysis of the bit error probability). We study the error exponents of these recently introduced bounding techniques. This work forms a direct continuation to the derivation of these bounds by Yousefi et al. [16, 17, 18] who also exemplified their superiority over the TSB for short binary linear block codes.

Putting the results reported by Divsalar [1] with the main result in this paper (see Theorem 1), we conclude that the error exponents of the simple bound of Divsalar [1], the first version of Duman and Salehi bounds [2], the TSB [9] and its improved versions by Yousefi et al. [15, 16, 17] all coincide. This conclusion holds for any ensemble of binary linear block codes (e.g., turbo codes, LDPC codes etc.) where we let the block lengths tend to infinity, so it does not only hold for the ensemble of fully random block codes (whose distance spectrum is binomially distributed). Moreover, the error exponents of the TSB versions for the bit error probability, as provided in [11, 19], coincide and are equal to the error exponent of the TSB for the block error probability. The explicit expression of this error exponent is given in Theorem 1, and is identical to the expression derived by Divsalar [1] for his simple bound. Based on Theorem 1, it follows that for any value of SNR, the same value of the normalized Hamming weight dominates the exponential behavior of the TSB and its two improved versions. In the asymptotic case where we let the block length tend to infinity, the dominating normalized Hamming weight can be explicitly calculated in terms of the SNR; this calculation is based on finding the value of the normalized Hamming weight $\delta$ which achieves the minimum in the RHS of (50), where this value clearly depends on the asymptotic growth rate of the distance spectrum of the ensemble under consideration. A similar calculation of this critical weight as a function of the SNR was done in [4], referring to the ensemble of fully random block codes and the simple union bound.

In a companion paper [14], new upper bounds on the block and bit error probabilities of linear block codes are derived. These bounds improve the tightness of the Shulman and Feder bound [13] and therefore also reproduce the random coding error exponent.

## Acknowledgment

The authors are grateful to the three anonymous reviewers for their constructive comments. The work was supported by the EU $6^{\text{th}}$ International Framework Programme via the NEWCOM Network of Excellence.

# Appendix

## A.1 The exponent of $\psi(\mathcal{C})$

In the following, the exponential behavior of the RHS of (48) is obtained by using the Chernoff bounding technique for $\psi(\mathcal{C})$.

Note that the geometrical region of the TSB corresponds to a double sided circular cone. For the derivation of the bound for the single cone, we have put the further restriction $z_1 \leq \sqrt{nE_{\rm s}}$, but since $z_1 \sim N(0, \frac{N_0}{2})$, then this boundary effect does not have any implication on the exponential behavior of the function $\psi(\mathcal{C})$ for large values of $n$ (as also noted in [1, p. 23]). To simplify the analysis, we therefore do not take into consideration of this boundary effect for large values of $n$. Let $\widetilde{\psi}(\mathcal{C})$ designate the function which is obtained by removing the event $z_1 \leq \sqrt{nE_{\rm s}}$ from the expression for $\psi(\mathcal{C})$ (see the RHS of (48)).

Let us designate the normalized Gaussian noise vector by $\nu$, i.e., $(\nu_1, \ldots, \nu_n) = \sqrt{\frac{2}{N_0}}(z_1, \ldots, z_n)$, and define $\eta \triangleq \tan^2\theta$. The Gaussian random vector has $n$ orthogonal components which are therefore statistically independent. From (4) and (42), the following equalities hold for BPSK modulated signals:

$$\begin{aligned}
r &= \sqrt{2nc\eta} \\
r_{\nu_1} &= \sqrt{\eta}\left(\sqrt{2nc} - \nu_1\right) \\
\beta_h(\nu_1) &= \left(\sqrt{2nc} - \nu_1\right)\sqrt{\frac{h}{n-h}} \\
l_{w,h}(\nu_1, \nu_2) &= \frac{\beta_w(\nu_1) - \rho_{w,h}\,\nu_2}{\sqrt{1-\rho_{w,h}^2}}.
\end{aligned} \tag{A.1}$$

Hence, we obtain from (48) and the above discussion

$$\begin{aligned}
\widetilde{\psi}(\mathcal{C}) = \min_w \Bigg\{ & \Pr\left(\sum_{i=2}^n \nu_i^2 \leq r_{\nu_1}^2, \nu_2 \geq \beta_w(\nu_1)\right) \\
& + \sum_{h=1}^n A_h \Pr\left(\sum_{i=2}^n \nu_i^2 \leq r_{\nu_1}^2, \nu_2 \geq \beta_h(\nu_1), \nu_3 \geq -l_{w,h}(\nu_1, \nu_2)\right) \\
& + \Pr\left(\sum_{i=2}^n \nu_i^2 \geq r_{\nu_1}^2\right)\Bigg\}.
\end{aligned} \tag{A.2}$$

At this point, we upper bound the RHS of (A.2) by the Chernoff bounds, namely, for three random variables $V, W$ and $Z$

$$\Pr(V \geq 0) \leq \mathbb{E}\left[e^{pV}\right], \quad p \geq 0 \tag{A.3}$$

$$\Pr(W \leq 0, V \geq 0) \leq \mathbb{E}\left[e^{qW+uV}\right], \quad q \leq 0, u \geq 0 \tag{A.4}$$

$$\Pr(W \leq 0, V \geq 0, Z \geq 0) \leq \mathbb{E}\left[e^{tW+sV+kZ}\right], \quad t \leq 0, s \geq 0, k \geq 0. \tag{A.5}$$

The Chernoff versions of the first and last terms in the RHS of (A.2) are introduced in [1, Eqs.(134)–(137)], and are given by

$$\Pr\left(\sum_{i=2}^{n}\nu_i^2 \geq r_{\nu_1}^2\right) \leq \sqrt{\frac{1-2p}{1+2p\eta}}e^{-nE_1(c,p,\eta)}, \quad p \geq 0 \tag{A.6}$$

$$\Pr\left(\sum_{i=2}^{n}\nu_i^2 \leq r_{\nu_1}^2, \nu_2 \geq \beta_w(\nu_1)\right) \leq \sqrt{\frac{1-2q}{1+2q\eta}}e^{-nE_2(c,q,\frac{w}{n},\eta)}, \quad -\frac{1}{2\eta} \leq q \leq 0 \tag{A.7}$$

where

$$E_1(c,p,\eta) = \frac{2p\eta c}{1+2p\eta} + \frac{1}{2}\ln(1-2p). \tag{A.8}$$

and

$$E_2(c,q,\delta,\eta) = c\left(\frac{2q\eta + (1-2q)\sqrt{\frac{\delta}{1-\delta}}}{1+2q\eta+(1-2q)\sqrt{\frac{\delta}{1-\delta}}}\right) + \frac{1}{2}\ln(1-2q). \tag{A.9}$$

Next, by invoking the Chernoff bound (A.5), we get an exponential upper bound on the second term in the RHS of (48). Using the notation

$$\zeta_{w,h} \triangleq \sqrt{\frac{w(n-h)}{h(n-w)}} \tag{A.10}$$

we get (see Appendix A.2 for details)

$$\begin{aligned} &A_h \Pr\left(\sum_{i=2}^{n}\nu_i^2 \leq r_{\nu_1}^2, \nu_2 \geq \beta_h(\nu_1), \nu_3 \geq -l_{w,h}(\nu_1,\nu_2)\right) \\ &\leq \sqrt{\frac{1-2t}{1+2t\eta}}e^{-g(c,t,k,s,\eta,h,n)}, \qquad -\frac{1}{2\eta} \leq t \leq 0,\ k \geq 0,\ s \geq 0 \end{aligned} \tag{A.11}$$

where

$$\begin{aligned} g(c,t,k,s,\eta,h,n) \triangleq\ & \frac{4t\eta nc + 2\sqrt{2nc}\left(s - \frac{k\zeta_{w,h}}{\sqrt{1-\rho_{w,h}^2}}\right)\Delta_h - \Delta_h^2\left(s - \frac{k\zeta_{w,h}}{\sqrt{1-\rho_{w,h}^2}}\right)^2}{2(1+2t\eta)} \\ & - \frac{\left(s - \frac{k\rho_{w,h}}{\sqrt{1-\rho_{w,h}^2}}\right)^2}{2(1-2t)} - \frac{k^2}{2(1-2t)} + \frac{n}{2}\ln(1-2t) - nr(\frac{h}{n}) \end{aligned} \tag{A.12}$$

and

$$\Delta_h \triangleq \sqrt{\frac{h}{n-h}}.$$

The next step is to find optimal values for $k$ and $s$ in order to maximize the function $g$. If $k^* = 0$ then the exponent of $\psi(\mathcal{C})$ is identical to that of the TSB. In order to find the optimal $k \geq 0$ and $s \geq 0$ which maximize $g$, we consider the aforementioned probabilities by discussing separately the three cases where $h < w$, $h > w$ and $h = w$.

*Case 1: $h = w$.* In this case $\zeta_{w,h} = \zeta_{w,w} = 1$, and we get

$$A_w \Pr\left(\sum_{i=2}^{n} \nu_i^2 \leq r_{\nu_1}^2, \nu_2 \geq \beta_w(\nu_1), \nu_3 \geq -l_{w,w}(\nu_1,\nu_2)\right) \leq \sqrt{\frac{1-2t}{1+2t\eta}} e^{-g(c,t,k,s,\eta,w,n)} \tag{A.13}$$

$$-\frac{1}{2\eta} \leq t \leq 0,\ k \geq 0,\ s \geq 0$$

where

$$\begin{aligned} g(c,t,k,s,\eta,w,n) \quad &= \quad \frac{4t\eta nc + 2\sqrt{2nc}\left(s - \frac{k}{\sqrt{1-\rho_{w,w}^2}}\right)\Delta_w - \Delta_w^2\left(s - \frac{k}{\sqrt{1-\rho_{w,w}^2}}\right)^2}{2(1+2t\eta)} \\ &- \quad \frac{\left(s - \frac{k\rho_{w,w}}{\sqrt{1-\rho_{w,w}^2}}\right)^2}{2(1-2t)} - \frac{k^2}{2(1-2t)} + \frac{n}{2}\ln(1-2t) - \ln(A_w). \end{aligned} \tag{A.14}$$

Let us define the parameters

$$\xi = s - \frac{k}{\sqrt{1-\rho_{w,w}^2}} \tag{A.15}$$

$$\tau = s - \frac{k\rho_{w,w}}{\sqrt{1-\rho_{w,w}^2}}. \tag{A.16}$$

From (A.15) and (A.16), we get

$$k = -(\xi - \tau)\alpha \tag{A.17}$$

where

$$\alpha \triangleq \sqrt{\frac{1+\rho_{w,w}}{1-\rho_{w,w}}}. \tag{A.18}$$

Hence, the Chernoff bounding technique gives

$$\Pr\left(\sum_{i=2}^{n} \nu_i^2 \leq r_{\nu_1}^2, \nu_2 \geq \beta_w(\nu_1), \nu_3 \geq -l_{w,w}(\nu_1,\nu_2)\right) \leq \sqrt{\frac{1-2t}{1+2t\eta}} e^{-g_1(c,t,\xi,\tau,\eta,w,n)} \tag{A.19}$$

$$-\frac{1}{2\eta} \leq t \leq 0$$

where

$$\begin{aligned} g_1(c,t,\xi,\tau,\eta,h,n) \quad &= \quad \frac{4t\eta nc + 2\sqrt{2nc}\xi\Delta_w - \Delta_w^2\xi^2}{2(1+2t\eta)} \\ &- \quad \frac{\tau^2}{2(1-2t)} - \frac{(\xi-\tau)^2\alpha^2}{2(1-2t)} + \frac{n}{2}\ln(1-2t). \end{aligned} \tag{A.20}$$

Maximizing the RHS of (A.19) w.r.t. $\tau$ yields

$$\begin{aligned} \frac{\partial g_1}{\partial \tau} \quad &= \quad -\frac{\tau}{1-2t} + \frac{(\xi-\tau)\alpha^2}{1-2t} = 0 \\ \Rightarrow \tau^* \quad &= \quad \frac{\alpha^2\xi^*}{1+\alpha^2}. \end{aligned} \tag{A.21}$$

Notice that $\frac{\partial^2 g_1}{\partial \tau^2} < 0$, hence plugging $\tau^*$ in (A.20) maximizes $g_1$. Substituting $\tau^*$ into (A.20) gives

$$\begin{aligned} g_2(c,t,\xi,\eta,w,n) &\triangleq g_1(c,t,\xi,\tau^*,\eta,w,n) \\ &= \frac{4t\eta nc + 2\sqrt{2nc}\Delta_w\xi - \Delta_w^2\xi^2}{2(1+2t\eta)} - \frac{\frac{\alpha^2}{1+\alpha^2}\xi^2}{2(1-2t)} + \frac{n}{2}\ln(1-2t). \end{aligned} \quad \text{(A.22)}$$

A differentiation of $g_2$ w.r.t. $\xi$ and an introduction of the new parameter $\epsilon \triangleq \frac{\alpha^2}{1+\alpha^2}$ gives

$$\begin{aligned} \frac{\partial g_2}{\partial \xi} &= \frac{\sqrt{2nc}\Delta_w - \Delta_w^2\xi}{1+2t\eta} - \frac{\epsilon\xi}{1-2t} = 0 \\ \xi^* &= \frac{\sqrt{2nc}\Delta_w(1-2t)}{\Delta_w^2(1-2t)+\epsilon(1+2t\eta)}. \end{aligned} \quad \text{(A.23)}$$

Again, $\frac{\partial^2 g_2}{\partial \xi^2} < 0$, so $\xi^*$ maximizes $g_2$. From (A.21), $\xi^* - \tau^* > 0$. Since $\alpha$ is non-negative, we get that $k^*$ in (A.17) is not-positive. But since from (A.11), $k \geq 0$, this yields that the optimal value of $k$ is equal to zero. From the Chernoff bound in (A.5), an optimality of $k$ when it is set to zero implies that asymptotically, as $n \to \infty$

$$\Pr\left(\sum_{i=2}^{n}\nu_i^2 \leq r_{\nu_1}^2, \nu_2 \geq \beta_w(\nu_1), \nu_3 \geq -l_{w,w}(\nu_1,\nu_2)\right) \doteq \Pr\left(\sum_{i=2}^{n}\nu_i^2 \leq r_{\nu_1}^2, \nu_2 \geq \beta_w(\nu_1)\right). \quad \text{(A.24)}$$

*Case 2: $h > w$.* In this case, from (40) it is obvious that $\rho_{w,h} = \sqrt{\frac{w(n-h)}{h(n-w)}}$. Hence, for this case, we get that $\rho_{w,h} = \zeta_{w,h}$. From (A.12)

$$\begin{aligned} g(c,t,k,s,\eta,h,n) &= \frac{4t\eta nc + 2\sqrt{2nc}\left(s - \frac{k\zeta_{w,h}}{\sqrt{1-\zeta_{w,h}^2}}\right)\Delta_h - \Delta_h^2\left(s - \frac{k\zeta_{w,h}}{\sqrt{1-\zeta_{w,h}^2}}\right)^2}{2(1+2t\eta)} \\ &- \frac{\left(s - \frac{k\zeta_{w,h}}{\sqrt{1-\zeta_{w,h}^2}}\right)^2}{2(1-2t)} - \frac{k^2}{2(1-2t)} + \frac{n}{2}\ln(1-2t) - nr(\frac{h}{n}). \end{aligned} \quad \text{(A.25)}$$

In the following, we introduce the parameters

$$\xi \triangleq s - \frac{k\zeta_{w,h}}{\sqrt{1-\zeta_{w,h}^2}} \quad \text{(A.26)}$$

$$\tau \triangleq k. \quad \text{(A.27)}$$

Optimization over $\tau$ yields $\tau^* = 0$, so $k^* = 0$, and asymptotically (as we let $n$ tend to infinity), one gets the following equality in terms of the exponential behaviors:

$$\Pr\left(\sum_{i=2}^{n}\nu_i^2 \leq r_{\nu_1}^2, \nu_2 \geq \beta_h(\nu_1), \nu_3 \geq -l_{w,h}(\nu_1,\nu_2)\right) \doteq \Pr\left(\sum_{i=2}^{n}\nu_i^2 \leq r_{\nu_1}^2, \nu_2 \geq \beta_h(\nu_1)\right). \quad \text{(A.28)}$$

*Case 3:* $h < w$. From (40), the values of $h$ approve that $\rho_{w,h} = \sqrt{\frac{h(n-w)}{w(n-h)}}$, so we get from (A.10) that $\rho_{w,h} < \zeta_{w,h}$. Define

$$\xi \triangleq s - \frac{k\zeta_{w,h}}{\sqrt{1-\rho_{w,h}^2}} \tag{A.29}$$

$$\tau \triangleq s - \frac{k\rho_{w,h}}{\sqrt{1-\rho_{w,h}^2}}. \tag{A.30}$$

From (A.29) and (A.30)

$$k = -(\xi - \tau)\alpha' \tag{A.31}$$

where

$$\alpha' \triangleq \frac{\sqrt{1-\rho_{w,h}^2}}{\zeta_{w,h} - \rho_{w,h}}. \tag{A.32}$$

Since in this case $\rho_{w,h} < \zeta_{w,h}$, then $\alpha' > 0$. Similarly to the arguments in case 1, we get again that the optimal value for $k$ is $k^* = 0$, which implies (A.28) in the limit where the block length tends to infinity.

## A.2 Derivation of the Chernoff Bound in (A.11) with the Function $g$ in (A.12)

Using the Chernoff bound (A.5) and defining

$$\Delta_w \triangleq \sqrt{\frac{w}{n-w}} \tag{A.33}$$

we get

$$\begin{aligned}
&\Pr\left(\sum_{i=2}^{n} \nu_i^2 \le r_{\nu_1}^2, \nu_2 \ge \beta_h(\nu_1), \nu_3 \ge -l_{w,h}(\nu_1,\nu_2)\right)\\
&\overset{(a)}{\le} \mathbb{E}\left[e^{t\left(\sum_{i=2}^n \nu_i^2 - r_{\nu_1}^2\right) + s(\nu_2 - \beta_h(\nu_1)) + k\left(\nu_3 + l_{w,h}(\nu_1,\nu_2)\right)}\right], \qquad t \le 0,\ s \ge 0,\ k \ge 0\\
&\overset{(b)}{=} \mathbb{E}\left[e^{t\left(\sum_{i=2}^n \nu_i^2 - \eta(\sqrt{2nc}-\nu_1)^2\right) + s\left(\nu_2 - \Delta_h(\sqrt{2nc}-\nu_1)\right) + k\left(\nu_3 + \frac{\Delta_w(\sqrt{2nc}-\nu_1) - \rho_{w,h}\nu_2}{\sqrt{1-\rho_{w,h}^2}}\right)}\right]\\
&= \mathbb{E}\left[e^{t\sum_{i=2}^n \nu_i^2 - t\eta\nu_1^2 - 2tn\eta c + 2\eta t\sqrt{2nc}\nu_1 + s\nu_2 - s\Delta_h\sqrt{2nc} + s\Delta_h\nu_1 + k\nu_3 + \frac{k\Delta_w\sqrt{2nc}}{\sqrt{1-\rho_{w,h}^2}} - \frac{k(\Delta_w\nu_1 + \rho_{w,h}\nu_2)}{\sqrt{1-\rho_{w,h}^2}}}\right]\\
&\overset{(c)}{=} \mathbb{E}\left[e^{t\sum_{i=4}^n \nu_i^2}\right] \mathbb{E}\left[e^{-t\eta\nu_1^2 + \left(2\eta t\sqrt{2nc} + s\Delta_h - \frac{k\Delta_w}{\sqrt{1-\rho_{w,h}^2}}\right)\nu_1}\right] \mathbb{E}\left[e^{t\nu_2^2 + \left(s - \frac{k\rho_{w,h}}{\sqrt{1-\rho_{w,h}^2}}\right)\nu_2}\right]\\
&\quad \cdot \mathbb{E}\left[e^{t\nu_3^2 + k\nu_3}\right] e^{-2tn\eta c - s\Delta_h\sqrt{2nc} + \frac{k\Delta_w\sqrt{2nc}}{\sqrt{1-\rho_{w,h}^2}}}.
\end{aligned} \tag{A.34}$$

where inequality (a) follows from the Chernoff bound (A.5), equality (b) follows from (A.1), and equality (c) follows from the statistical independence of the components of the normalized noise vector $\nu$. For a zero-mean and unit-variance Gaussian random variable $X$, the following equality holds:

$$\mathbb{E}\left[e^{aX^2+bX}\right] = \frac{e^{\frac{b^2}{2(1-2a)}}}{\sqrt{1-2a}}, \quad a \leq \frac{1}{2},\ b \in \mathbb{R}. \tag{A.35}$$

Evaluating each term in (A.34) with the equality in (A.35), and substituting

$$\zeta_{w,h} = \frac{\Delta_w}{\Delta_h} \tag{A.36}$$

which follows from (A.10) and (A.33), then gives

$$\mathbb{E}\left[e^{t\sum_{i=4}^{n}\nu_i^2}\right] = \left(\frac{1}{\sqrt{1-2t}}\right)^{n-3}, \quad t \leq 0 \tag{A.37}$$

$$\mathbb{E}\left[e^{-t\eta\nu_1^2+\left(2\eta t\sqrt{2nc}+s\Delta_h-\frac{k\Delta_w}{\sqrt{1-\rho_{w,h}^2}}\right)\nu_1}\right] = \frac{1}{\sqrt{1+2t\eta}}\, e^{\frac{\left(2\eta t\sqrt{2nc}+\Delta_h\left(s-\frac{k\zeta_{w,h}}{\sqrt{1-\rho_{w,h}^2}}\right)\right)^2}{2(1+2t\eta)}} \tag{A.38}$$

$$\mathbb{E}\left[e^{t\nu_2^2+\left(s-\frac{k\rho_{w,h}}{\sqrt{1-\rho_{w,h}^2}}\right)\nu_2}\right] = \frac{1}{\sqrt{1-2t}}\, e^{\frac{\left(s-\frac{k\rho_{w,h}}{\sqrt{1-\rho_{w,h}^2}}\right)^2}{2(1-2t)}}, \quad k \geq 0,\ s \geq 0 \tag{A.39}$$

$$\mathbb{E}\left[e^{t\nu_3^2+k\nu_3}\right] = \frac{1}{\sqrt{1-2t}}\, e^{\frac{k^2}{2(1-2t)}}, \quad t \leq 0,\ k \geq 0. \tag{A.40}$$

From (A.38), straightforward algebra gives

$$\mathbb{E}\left[e^{-t\eta\nu_1^2+\left(2\eta t\sqrt{2nc}+s\Delta_h-\frac{k\Delta_w}{\sqrt{1-\rho_{w,h}^2}}\right)\nu_1}\right] e^{-2tn\eta c-s\Delta_h\sqrt{2nc}+\frac{k\Delta_w\sqrt{2nc}}{\sqrt{1-\rho_{w,h}^2}}}$$

$$= \frac{1}{\sqrt{1+2t\eta}} \exp\left\{\frac{-4t\eta nc - 2\sqrt{2nc}\left(s-\frac{k\zeta_{w,h}}{\sqrt{1-\rho_{w,h}^2}}\right)\Delta_h + \Delta_h^2\left(s-\frac{k\zeta_{w,h}}{\sqrt{1-\rho_{w,h}^2}}\right)^2}{2(1+2t\eta)}\right\}. \tag{A.41}$$

Plugging (A.37) and (A.39)–(A.41) into (A.34) finally gives

$$
\begin{aligned}
&A_h \Pr\left(\sum_{i=2}^{n} \nu_i^2 \le r_{\nu_1}^2, \nu_2 \ge \beta_h(\nu_1), \nu_3 \ge -l_{w,h}(\nu_1,\nu_2)\right)\\
&\le \frac{A_h}{\sqrt{1+2t\eta}}\left(\frac{1}{\sqrt{1-2t}}\right)^{n-1} e^{\frac{-4t\eta nc-2\sqrt{2nc}\left(s-\frac{k\zeta_{w,h}}{\sqrt{1-\rho_{w,h}^2}}\right)\Delta_h+\Delta_h^2\left(s-\frac{k\zeta_{w,h}}{\sqrt{1-\rho_{w,h}^2}}\right)^2}{2(1+2t\eta)}+\frac{\left(s-\frac{k\rho_{w,h}}{\sqrt{1-\rho_{w,h}^2}}\right)^2}{2(1-2t)}+\frac{k^2}{2(1-2t)}}\\
&= \sqrt{\frac{1-2t}{1+2t\eta}}\, e^{-g(c,t,k,s,\eta,h,n)}, \qquad -\frac{1}{2\eta} < t \le 0,\ k \ge 0,\ s \ge 0
\end{aligned}
\tag{A.42}
$$

which proves the Chernoff bound in (A.11) with the function $g$ introduced in (A.12).

### A.3 Monotonicity w.r.t. the Correlation Coefficient

Consider the probabilities $\Pr(E_{0\to i}, E^{\mathrm{c}}_{0\to j}, \mathbf{y} \in C_n(\theta)|z_1)$, and denote the Hamming weights of $\mathbf{c}_i$ and $\mathbf{c}_j$ by $d_i$ and $d_j$, respectively. In [17], it is shown that as long as $d_i > d_j$, the probabilities $\Pr(E_{0\to i}, E^{\mathrm{c}}_{0\to j}, \mathbf{y} \in C_n(\theta)|z_1)$ are monotonically decreasing functions of the correlation coefficients $\rho$ between the planes $(\mathbf{o}, \mathbf{s}_0, \mathbf{s}_i)$ and $(\mathbf{o}, \mathbf{s}_0, \mathbf{s}_j)$. Hence, the complex optimization problem in (24) is simplified by choosing the first error event as well as the complementary error events in the RHS of (24) to correspond to a codeword with Hamming weight $d_{\min}$, and (25) is obtained. Here we prove, that the aforementioned probabilities are monotonically decreasing functions of the correlation coefficients for *any* choice of $i, j$. As a consequence, one can obtain a version of the ITSB by setting in (24) $\pi_1 = \lambda_i = w$ where $w \in \{d_{\min}, \ldots, d_{\max}\}$, and choosing the optimal $w$ which minimizes the resulting upper bound. In order to prove this, we follow the steps in [17, Appendix I] where it is shown that the above probabilities are monotonically decreasing functions of $\rho$ if

$$\frac{z_2}{\beta_j(z_1)} > \rho. \tag{A.43}$$

Note that the joint event $(E_{0\to i},\ \mathbf{y} \in C_n(\theta))$ implies that the noise component $z_2$ is in the range between $\beta_i(z_1)$ and $r_{z_1}$ (see Fig. 1), so the minimum value of the RHS of (A.43) is

$$\frac{\beta_i(z_1)}{\beta_j(z_1)} = \sqrt{\frac{d_i(n-d_j)}{d_j(n-d_i)}}.$$

Clearly,

$$\sqrt{\frac{d_i(n-d_j)}{d_j(n-d_i)}} > \frac{\min(d_i,d_j)[n-\max(d_i,d_j)]}{\sqrt{d_i d_j (n-d_i)(n-d_j)}} \tag{A.44}$$

but from (34), it is evident that the RHS of (A.44) is the maximal value of $\rho$, thus, condition (A.43) is always satisfied referring to the joint event $(E_{0\to i},\ \mathbf{y} \in C_n(\theta))$.

## References

[1] D. Divsalar, "A simple tight bound on error probability of block codes with application to Turbo codes," *TMO progress Report 42-139* NASA, JPL, Pasadena, CA, USA, 1999.

[2] T. M. Duman and M. Salehi, "New performance bounds for turbo codes," *IEEE Trans. on Communications*, vol. 46, pp. 717–723, June 1998.

[3] T. M. Duman, *Turbo Codes and Turbo Coded Modulation Systems: Analysis and Performance Bounds*, Ph.D. dissertation, Elect. Comput. Eng. Dep., Northeastern University, Boston, MA, USA, May 1998.

[4] M. Fossorier, "Critical point for maximum-likelihood decoding of linear block codes," *IEEE Communications Letters*, vol. 9, no. 9, pp. 817–819, September 2005.

[5] J. Galambos and I. Simonelli, *Bonferroni-type inequalities with applications*, Springer Series in Statistics, Probability and its Applications, Springer-Verlag, New-York, 1996.

[6] R. G. Gallager, "A simple derivation of the coding theorem and some applications," *IEEE Trans. on Information Theory*, vol. 11, pp. 3–18, January 1965.

[7] H. Herzberg and G. Poltyrev, "Techniques of bounding the probability of decoding error for block modulation structures," *IEEE trans. on Information Theory,* vol. 40, no. 3, pp. 903–911, May 1994.

[8] D. Hunter, "An upper bound for the probability of a union,“ *Journal of Applied Probability*, vol. 13, pp. 597–603, 1976.

[9] G. Poltyrev, "Bounds on the decoding error probability of binary linear codes via their spectra," *IEEE Trans. on Information Theory*, vol. 40, no. 4, pp. 1284–1292, July 1994.

[10] I. Sason and S. Shamai, "Bounds on the error probability of ML decoding for block and turbo-block codes," *Annals of Telecommunication*, vol. 54, no. 3–4, pp. 183–200, March–April 1999.

[11] I. Sason and S. Shamai, "Improved upper bounds on the ML decoding error probability of parallel and serial concatenated turbo codes via their ensemble distance spectrum," *IEEE Trans. on Information Theory*, vol. 46, no. 1, pp. 24–47, January 2000.

[12] I. Sason and S. Shamai, "Performance analysis of linear codes under maximum-likelihood decoding: a tutorial," *Foundations and Trends in Communications and Information Theory*, vol. 3, no. 1–2, pp. 1–222, NOW Publishers, Delft, the Netherlands, July 2006.

[13] N. Shulman and M. Feder, "Random coding techniques for nonrandom codes,"*IEEE Trans. on Information Theory*, vol. 45, no. 6, pp. 2101–2104, September 1999.

[14] M. Twitto, I. Sason and S. Shamai, "Tightened upper bounds on the ML decoding error probability of binary linear block codes," submitted to the *IEEE Trans. on Information Theory*, February 2006. [Online]. Available: `http://arxiv.org/abs/cs.IT/0607003`.

[15] S. Yousefi and A. Khandani, "Generelized tangential-sphere bound on the ML decoding error probability of linear binary block codes in AWGN interference," *IEEE Trans. on Information Theory*, vol. 50, no. 11, pp. 2810–2815, November 2004.

[16] S. Yousefi and A. K. Khandani, "A new upper bound on the ML decoding error probability of linear binary block codes in AWGN interference," *IEEE Trans. on Information Theory*, vol. 50, no. 12, pp. 3026–3036, December 2004.

[17] S. Yousefi and A. Mehrabian, "Improved tangential-sphere bound on the ML decoding error probability of linear binary block codes in AWGN interference," *Proceedings* $37^{th}$ *Annual Conference on Information Science and Systems (CISS 2005),* John Hopkins University, Baltimor, MD, USA, March 16–18, 2005.

[18] S. Yousefi, "Gallager first bounding technique for the performance evaluation of maximum-likelihood decoded linear binary block codes," *IEE Proceedings on Communications*, vol. 153, no. 3, pp. 317–332, June 2006.

[19] J. Zangl and R. Herzog, "Improved tangential-sphere bound on the bit error probability of concatenated codes," *IEEE Journal on Selected Areas in Communications*, vol. 19, no. 5, pp. 825–837, May 2001.

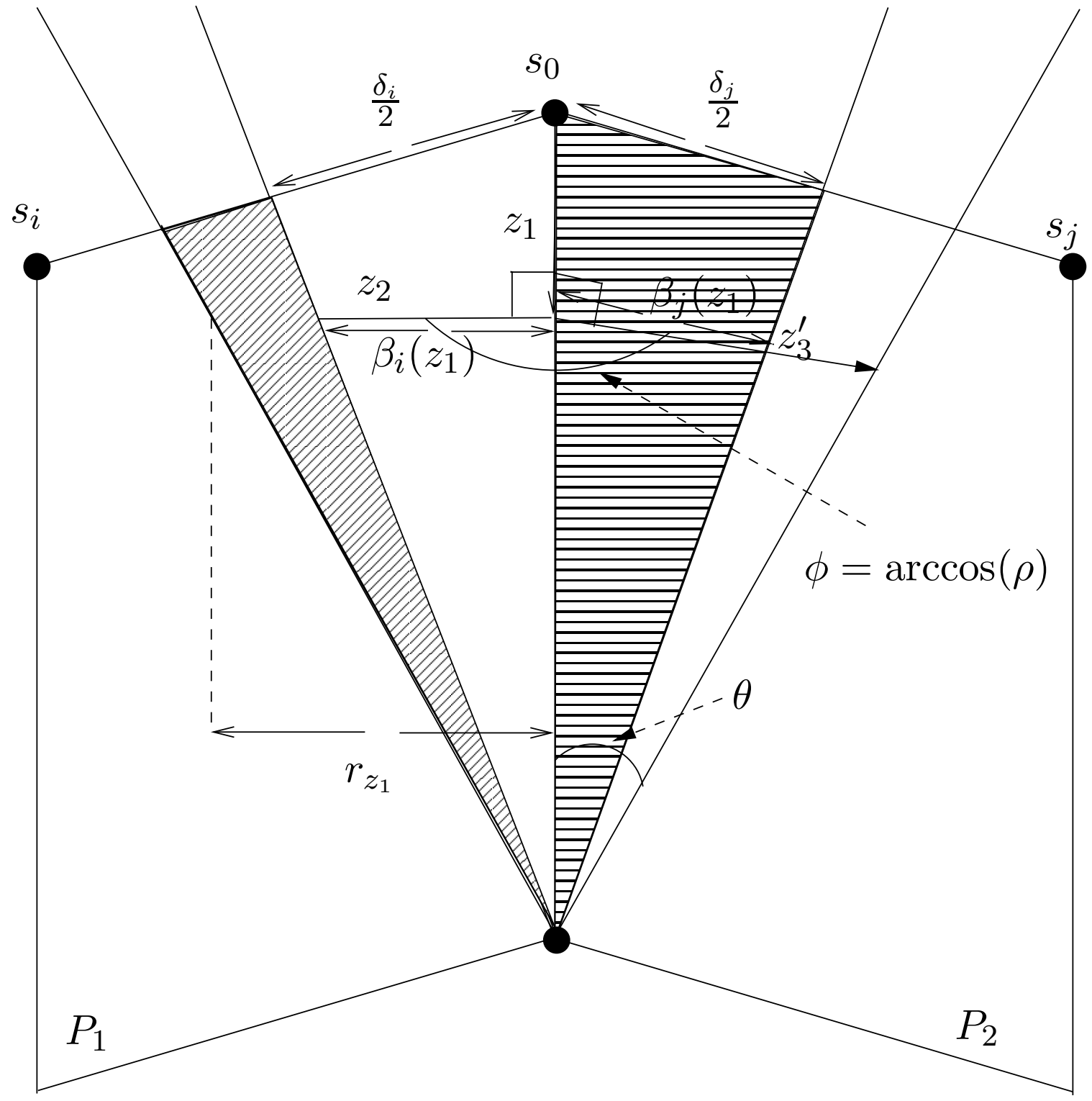

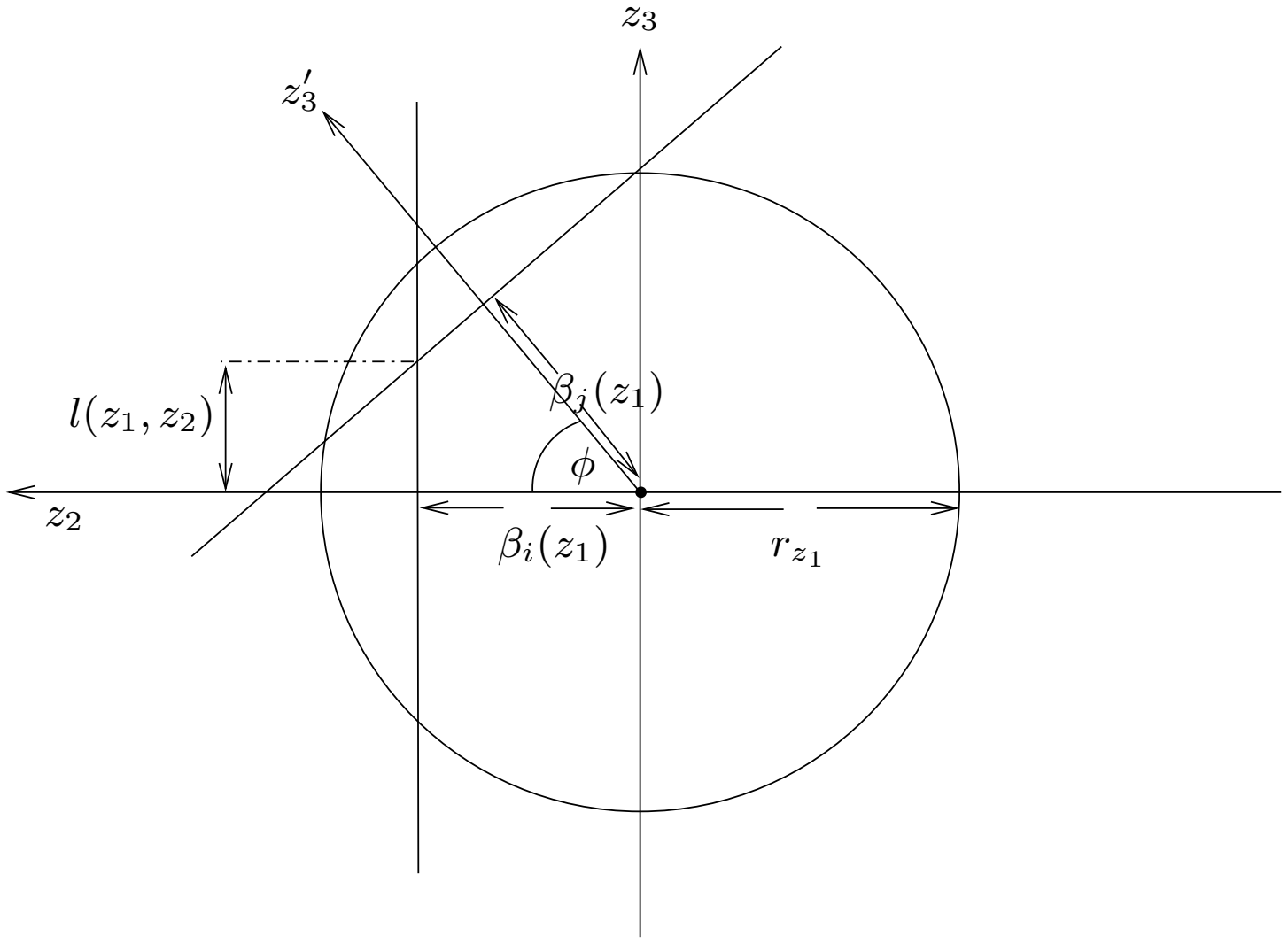

Figure 2: (a): $\mathbf{s}_0$ is the transmitted vector, $z_1$ is the radial noise component, $z_2$ and $z_3'$ are two (not necessarily orthogonal) noise components, which are perpendicular to $z_1$, and lie on planes $P_1$ and $P_2$, respectively. The doted and dashed areas are the regions where $E_i$ and $E_j^c$ occur, respectively. (b): A cross-section of the geometry in (a).

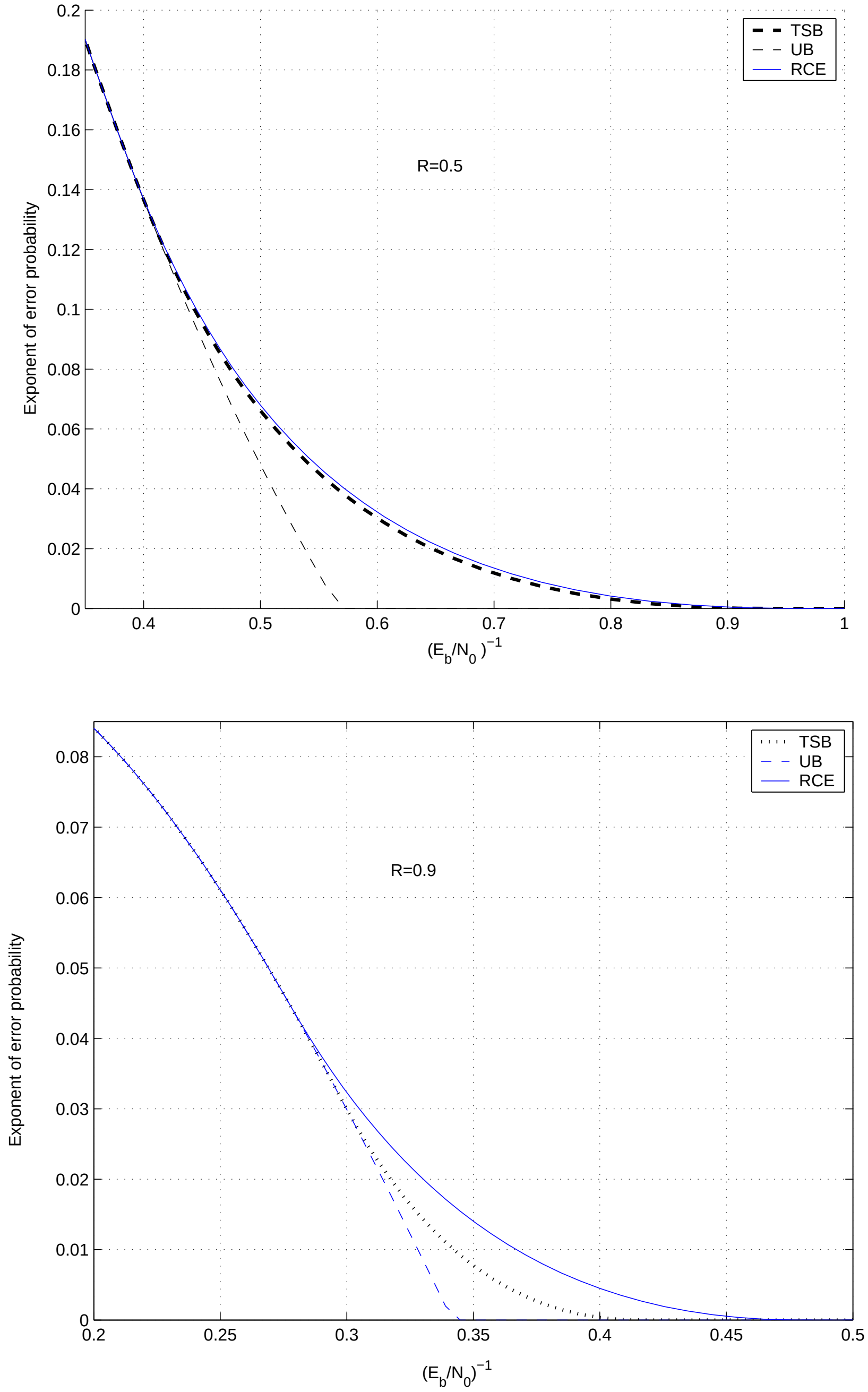

Figure 3: Comparison between the error exponents for random block codes which are based on the union bound (UB), the tangential-sphere bound (TSB) of Poltyrev [9] (which according to Theorem 1 is identical to the error exponents of the ITSB and the AHP bounds), and the random coding bound (RCE) of Gallager [6]. The upper and lower plots refer to code rates of 0.5 and 0.9 bits per channel use, respectively. The error exponents are plotted versus the reciprocal of the energy per bit to the one-sided spectral noise density.